\documentclass[aps,pre,showpacs,twocolumn]{revtex4}
\usepackage{graphicx}
\usepackage{dcolumn}
\usepackage{bm}

\expandafter\let\csname equation*\endcsname\relax
\expandafter\let\csname endequation*\endcsname\relax
\usepackage{amsmath,amssymb,amsthm}
\usepackage{epsfig}
\usepackage{graphicx}
\usepackage{array,color}
\usepackage{epstopdf}

\newcommand{\be}[0]{\begin{equation}}
\newcommand{\ee}[0]{\end{equation}}
\newcommand{\bee}[0]{\begin{equation*}}
\newcommand{\eee}[0]{\end{equation*}}
\newcommand{\bal}[0]{\begin{aligned}}
\newcommand{\eal}[0]{\end{aligned}}
\begin{document}

\preprint{APS/123-QED}

\title{Interactions of soliton and mean field in KdV equation with well type initial data}

\author{Ruizhi Gong}
\author{Deng-Shan Wang}%
 \email{dswang@bnu.edu.cn}
\affiliation{%
 Laboratory of Mathematics and Complex Systems (Ministry of Education),  \\ School of Mathematical Sciences, Beijing Normal University, Beijing 100875, China.
}%




\date{\today}

\begin{abstract}
For the KdV equation with well-type initial value, the interaction between the trial soliton and the mean field is studied. The well initial value will lead to the appearance of rarefaction wave and dispersion shock wave, and there will be a linear wave region after a long time. The interaction between trial soliton and mean field is described within the framework of Whitham modulation theory, and the trajectory of soliton is given. The predicted soliton amplitude and phase changes are numerically confirmed, verifying the correctness of the theoretical analysis.
\end{abstract}

\maketitle


\section{Introduction}\label{sec:1}
The interactions between nonlinear waves and mean field have attracted widespread attention as a
common problem in fluid mechanics \cite{Zabusky-1965}. This paper focuses on the interactions between solitary wave and
mean field (no distinction is made here between soliton and solitary wave). The key to the interactions
between small-scale dispersive wave and large-scale mean field is scale separation, which requires that
the characteristic length and time scale of the soliton be much smaller than the characteristic length and
time scale of the mean field in the problem we are considering \cite{EL-PRE-2018}-\cite{EL-JFM-2021}. The
KdV equation, as a classical equation of an integrable system \cite{Lax-1968,Gardner-1967} and a representative of the shallow water wave
model, is suitable for the study of soliton-mean field problem. Consider the KdV equation in classical form
\cite{Miura-1976}
\begin{equation}
\label{KdV}
u_t+6uu_x+u_{xxx}=0
\end{equation}
with the initial data that this paper focuses on are as follows:
\begin{equation}
\label{Initial condition}
u(x,0)=u_0=U+s_0,
\end{equation}
where
\begin{equation}
\label{trial}
s_0=s(x,0;x_0)=a\mathrm{sech}^2\left(\sqrt{\frac{a}{2}}(x-x_0)\right),
\end{equation}
and
\begin{equation}
\label{well}
U=\left\{
\begin{aligned}
&U_0<0,  &0<x<l,\\
&0,  &{\rm other ~else}.\\
\end{aligned}
\right.
\end{equation}
Observe that the initial value consists of a rectangular well (\ref{well}) and a stationary soliton solution of the exact form (\ref{trial}). The present work labels this initial soliton as the trial soliton. The well potential here forms a slowly varying mean field after time evolution, which is composed of rarefaction wave (RW), dispersive shock wave (DSW), plateau and linear wave (LW). Unlike the traditional well potential, the well here changes slowly with time, and the soliton follows the same evolution equation as the mean field. This phenomenon, which is different from the traditional interaction between nonlinear waves and mean fields, was first discovered by Maiden et al. \cite{EL-PRL-2018} in fluid dynamics experiments, and there are two possibilities: the solitary wave completely passes through the mean field and propagates with a new amplitude and speed, or it is embedded in the mean field and propagates cooperatively. Recently, Congy et al. \cite{EL-JFM-2019} completed the interaction of linear modulated waves with nonlinear, large-scale fluid states, and Sande et al. \cite{EL-JFM-2021} extended the interaction between soliton and mean field to non-convex flow.
\par
For the interaction between such finite amplitude wave and the mean field, Whitham modulation theory can be naturally applied as the basic framework \cite{Whitham-1965}-\cite{Luo-stud-2023}. It is worth pointing out that the soliton and the mean field satisfy the same modulation equation. It is clear here that the distance between the soliton and the discontinuity is much larger than the soliton width $L_S$ and the mean field shows a width $L_M$ that expands with time, satisfying $L_M\gg L_S$. This is the standard assumption of Whitham modulation theory to describe this multi-scale structure \cite{Whitham-1974}.
\par
The paper is organized as follows. In Section \ref{sec:2}, the Whitham modulation theory of the KdV equation is briefly reviewed, including the modulated periodic solutions and the modulation systems that they satisfy. The boundaries of each region in the $(x,t)$ plane obtained by the mean field evolution are calculated. In Section \ref{sec:3}, based on the modulation system derived by Whitham average method, the Riemann invariants of the soliton modulation system are derived. The transmission condition and phase relation of the trial soliton with the initial value of the well are given. The section \ref{sec:4}, as the main part of this paper, discusses the interactions between soliton and mean field when the trial soliton is placed in different positions (left, middle, and right). In view of different phenomena, the Riemann invariants are constructed and analyzed within the framework of Whitham modulation theory. Predictions of the soliton trajectories in various regions are given and the theoretical results are compared with the numerical results, showing good agreement. Some conclusions are summarized in Section \ref{sec:5}.

\section{Whitham modulation theory}\label{sec:2}

In this section, we briefly describe some essential results, most of which rely on the groundbreaking work done in 1980 \cite{FFM-CPAM-1980} and its extension.

\subsection{Periodic modulation solution and KdV-Whitham system}
As usual, take $\xi=x-Vt$ to obtain the traveling wave solution of the KdV equation. The detailed derivation process is not shown here, see \cite{EL-WIT}-\cite{arXiv-2024}. The periodic solution of KdV equation can be expressed as Riemann invariant and written as
\begin{equation}
\label{solution}
u=\lambda_1-\lambda_2+\lambda_3+2(\lambda_2-\lambda_1)\mathrm{cn}^2(2\sqrt{(\lambda_3-\lambda_1)}(x-Vt)+\xi_0,m),
\end{equation}
where $m=\frac{\lambda_2-\lambda_1}{\lambda_3-\lambda_1}$, $V=2(\lambda_1+\lambda_2+\lambda_3)$ and $\xi_0$ is an arbitrary phase position. The order relation of the Riemann invariants is specified as $\lambda_3\geq \lambda_2 \geq \lambda_1$.
The mean of the modulation period solution (\ref{solution}) is
\begin{equation}
\label{mean}
\left<u\right>=\frac{1}{2\pi}\int_0^{2\pi}u(\xi)d\xi=\lambda_1+\lambda_2-\lambda_3+2(\lambda_3-\lambda_1)\frac{E(m)}{K(m)}.
\end{equation}
As a preliminary work, the limit states also need to be expressed. The harmonic limit $\lambda_2\rightarrow\lambda_1$ corresponds to small amplitude state
\begin{equation}
u\sim\lambda_3+2(\lambda_2-\lambda_1)\mathrm{cos}^2(2\sqrt{(\lambda_3-\lambda_1)}\xi+\xi_0),
\end{equation}
and the soliton limit $\lambda_2\rightarrow\lambda_3$ corresponds to soliton state
\begin{equation}
u=\lambda_1+2(\lambda_{23}-\lambda_1)\mathrm{sech}^2(2\sqrt{(\lambda_{23}-\lambda_1)}(x-(2\lambda_1+4\lambda_{23})t)).
\end{equation}

The one-phase KdV-Whitham modulation system is allowed to be shown in a diagonal form
\begin{equation}\label{KdV-Whitham-E}
\frac{\partial \lambda_j}{\partial t}+v_j(\lambda)\frac{\partial \lambda_j}{\partial x}=0,\quad ~j=1,2,3,
\end{equation}
where the Whitham velocities $v_j~(j=1,2,3)$ are given in terms of group velocity
\begin{equation}
\label{vj}
v_j=\frac{\partial_j \omega}{\partial_j k}=V-2\frac{L}{\partial_j L},\quad ~j=1,2,3,
\end{equation}
where $\omega$ is the frequency and $k$ is the wave number. And the wavelength $L$ can be written by using Riemann invariants
\begin{equation}
L=\frac{2K(m)}{\sqrt{\lambda_3-\lambda_1}}.
\end{equation}
Specifically, we have
\begin{equation}
\begin{aligned}
&v_1=2(\lambda_1+\lambda_2+\lambda_3)-4(\lambda_2-\lambda_1)\frac{K(m)}{K(m)-E(m)},\\
&v_2=2(\lambda_1+\lambda_2+\lambda_3)-4(\lambda_2-\lambda_1)\frac{(1-m)K(m)}{E(m)-(1-m)K(m)},\\
&v_3=2(\lambda_1+\lambda_2+\lambda_3)+4(\lambda_3-\lambda_2)\frac{K(m)}{E(m)},
\end{aligned}
\end{equation}
where the modulus $m=\frac{\lambda_2-\lambda_1}{\lambda_3-\lambda_1}$ $(0\leq m\leq 1)$, $K(m)$ and $E(m)$ are complete elliptic integrals of the first and second kind, respectively.
Here we use the asymptotic expansions of the complete elliptic integrals
$$
\begin{aligned}
m\ll 1:\quad &K(m)=\frac{\Pi}{2}(1+\frac{m}{4}+\frac{9m^2}{64}+\ldots),\\
 &E(m)=\frac{\Pi}{2}(1-\frac{m}{4}-\frac{3m^2}{64}+\ldots),\\
(1-m)\ll 1:\quad &K(m)\approx \frac{1}{2}ln\frac{16}{1-m},\\
 &E(m)\approx 1+\frac{1}{4}(1-m)(ln\frac{16}{1-m}-1).
\end{aligned}
$$
Two asymptotic limits can be identified. In the harmonic limit $m\rightarrow 0$ (i.e., $\lambda_2\rightarrow \lambda_1$), the Whitham system degenerates into
\begin{equation}
\begin{aligned}
&\frac{\partial \lambda_{12}}{\partial t}+(12\lambda_{12}-\lambda_{3})\frac{\partial \lambda_{12}}{\partial x}=0,\\
&\frac{\partial \lambda_{3}}{\partial t}+6\lambda_{3}\frac{\partial \lambda_{3}}{\partial x}=0.
\end{aligned}
\end{equation}
In the soliton limit $m\rightarrow 1$ (i.e., $\lambda_2\rightarrow \lambda_3$), the Whitham system reduces to
\begin{equation}
\begin{aligned}
&\frac{\partial \lambda_{1}}{\partial t}+6\lambda_{1}\frac{\partial \lambda_{1}}{\partial x}=0,\\
&\frac{\partial \lambda_{23}}{\partial t}+(2\lambda_{1}+4\lambda_{23})\frac{\partial \lambda_{23}}{\partial x}=0.
\end{aligned}
\end{equation}
Based on the pioneering work of Flaschka, Forest, and McLaughlin \cite{FFM-CPAM-1980}, the Whitham modulation system describing the $N$-phase solution of the KdV equation can be obtained. The results of the one-phase have been given before. As preliminary work, the $N=2$ genus Whitham modulation equation is written below
\begin{equation}\label{N2-Whitham}
\frac{\partial \lambda_k}{\partial t}+v_k^{(2)}\frac{\partial \lambda_k}{\partial x}=0,\quad k=1,2,\cdots ,5.
\end{equation}
Introduce the following integral representation
\begin{equation}
I_i^j(\lambda_k)=\int_{\lambda_{2i-1}}^{\lambda_{2i}}\frac{(\lambda_k-\mu)\mu^j d\mu}{P(\mu)},
\end{equation}
where $P(\mu)=\sqrt{(\mu-\lambda_1)(\mu-\lambda_2)(\mu-\lambda_3)(\mu-\lambda_4)(\mu-\lambda_5)}$, then the Whitham velocities in (\ref{N2-Whitham}) can be expressed as
\begin{equation}
v_k^{(2)}=-6\Sigma_{k=1}^5 \lambda_k+12\lambda_k+12\frac{I_2^2(\lambda_k)I_1^0(\lambda_k)-I_2^0(\lambda_k)I_1^2(\lambda_k)}
{I_2^1(\lambda_k)I_1^0(\lambda_k)-I_2^0(\lambda_k)I_1^1(\lambda_k)},
\end{equation}
for $k=1,2,\cdots ,5.$

\subsection{Formation of linear wave train}
\begin{figure*}
\centering
\includegraphics[width=7.0cm]{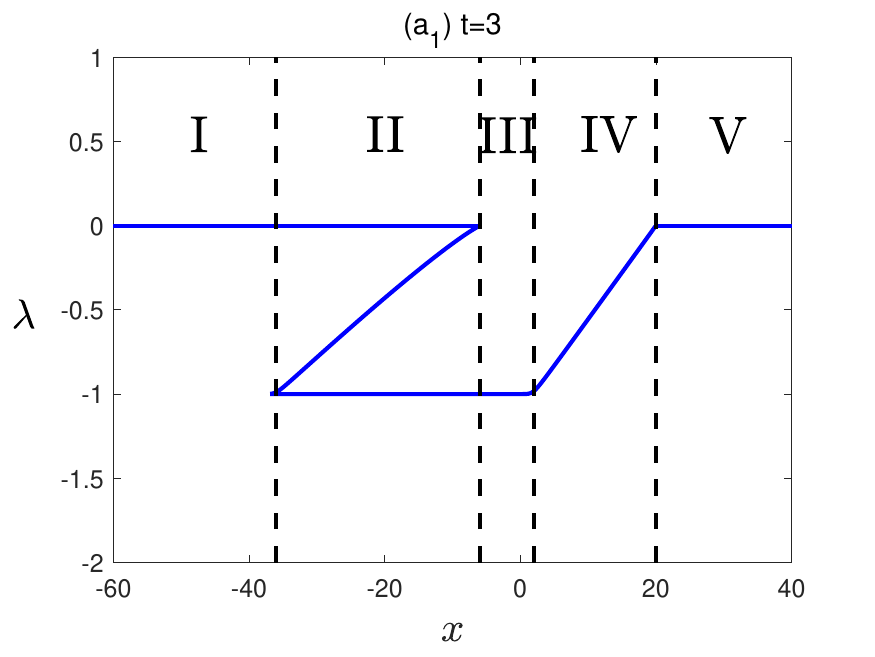}
\includegraphics[width=7.0cm]{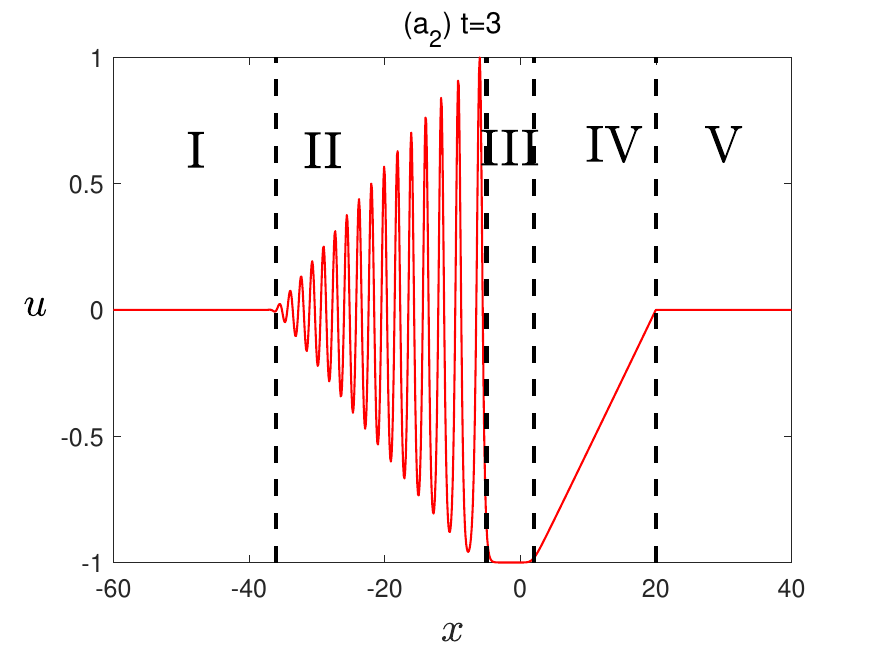}
\includegraphics[width=7.0cm]{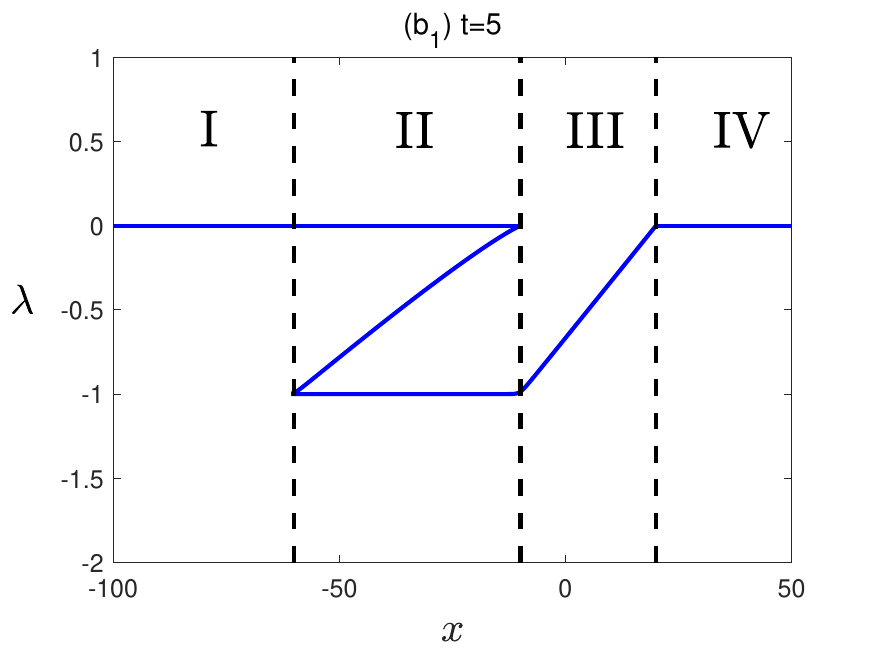}
\includegraphics[width=7.0cm]{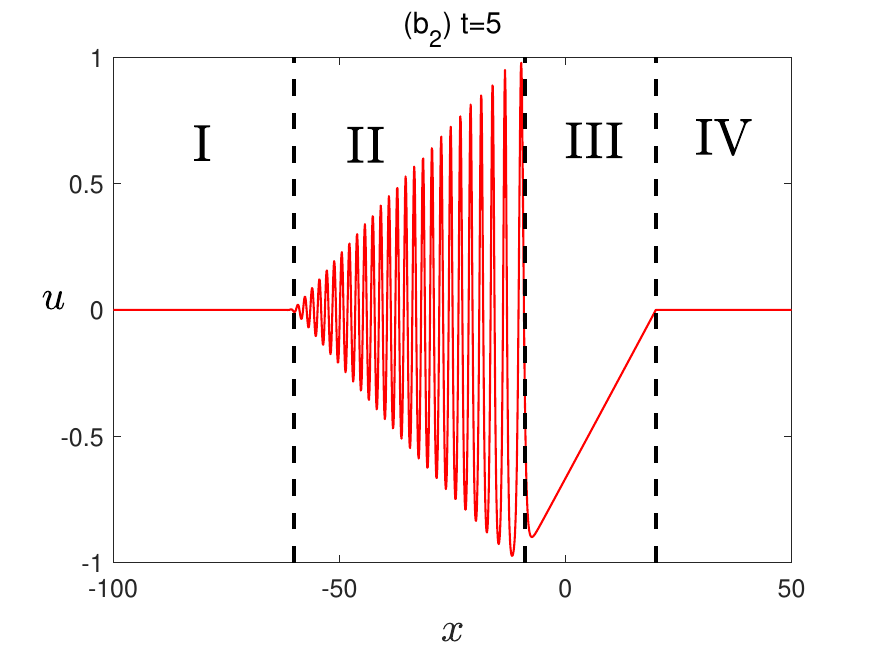}
\includegraphics[width=7.0cm]{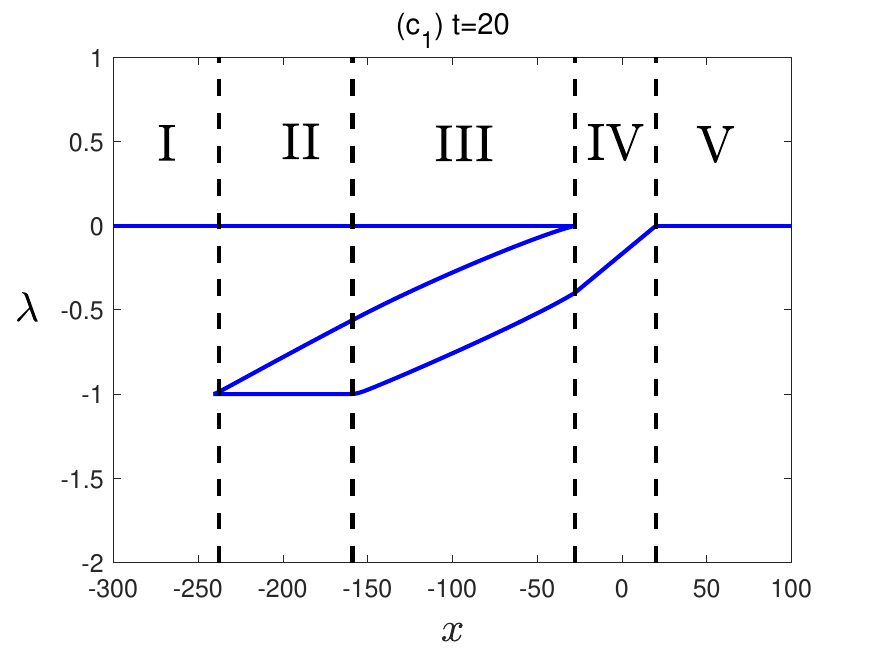}
\includegraphics[width=7.0cm]{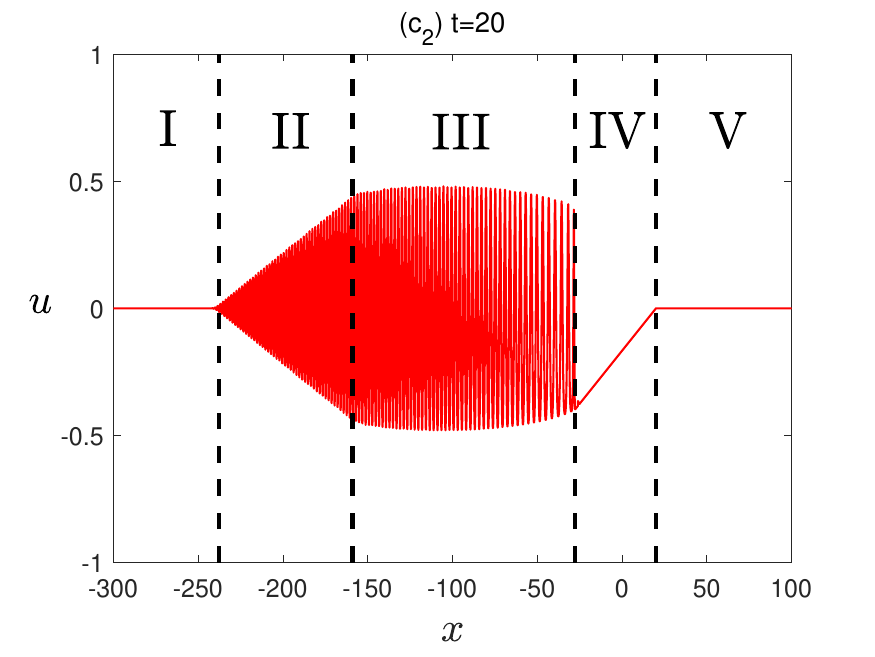}
\includegraphics[width=7.0cm]{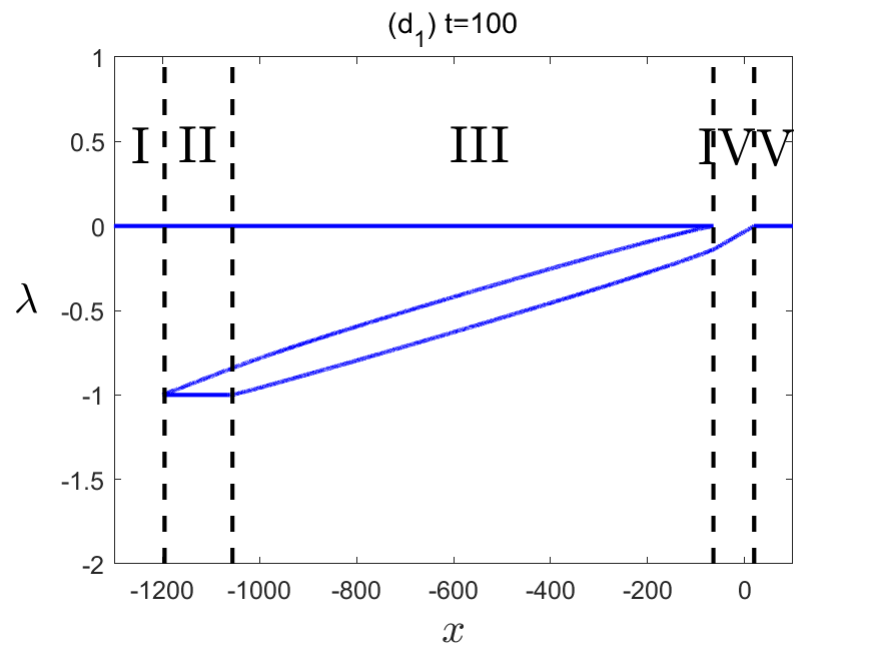}
\includegraphics[width=7.0cm]{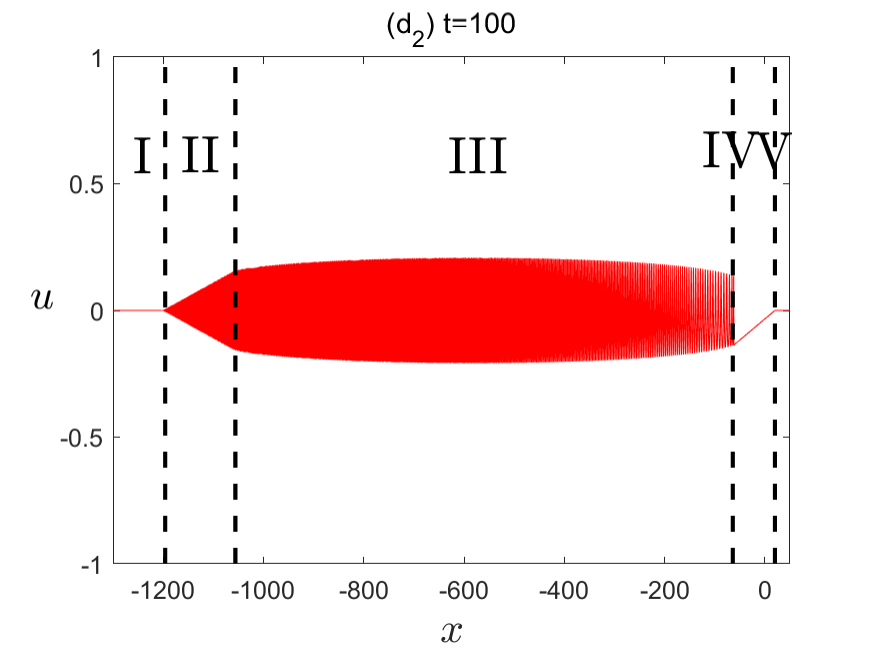}
\caption{{\protect\small Long-term evolution of the mean field of the well initial value.}}
\label{lambda-whitham-fig}
\end{figure*}

\begin{figure}
\centering
\includegraphics[width=8.0cm]{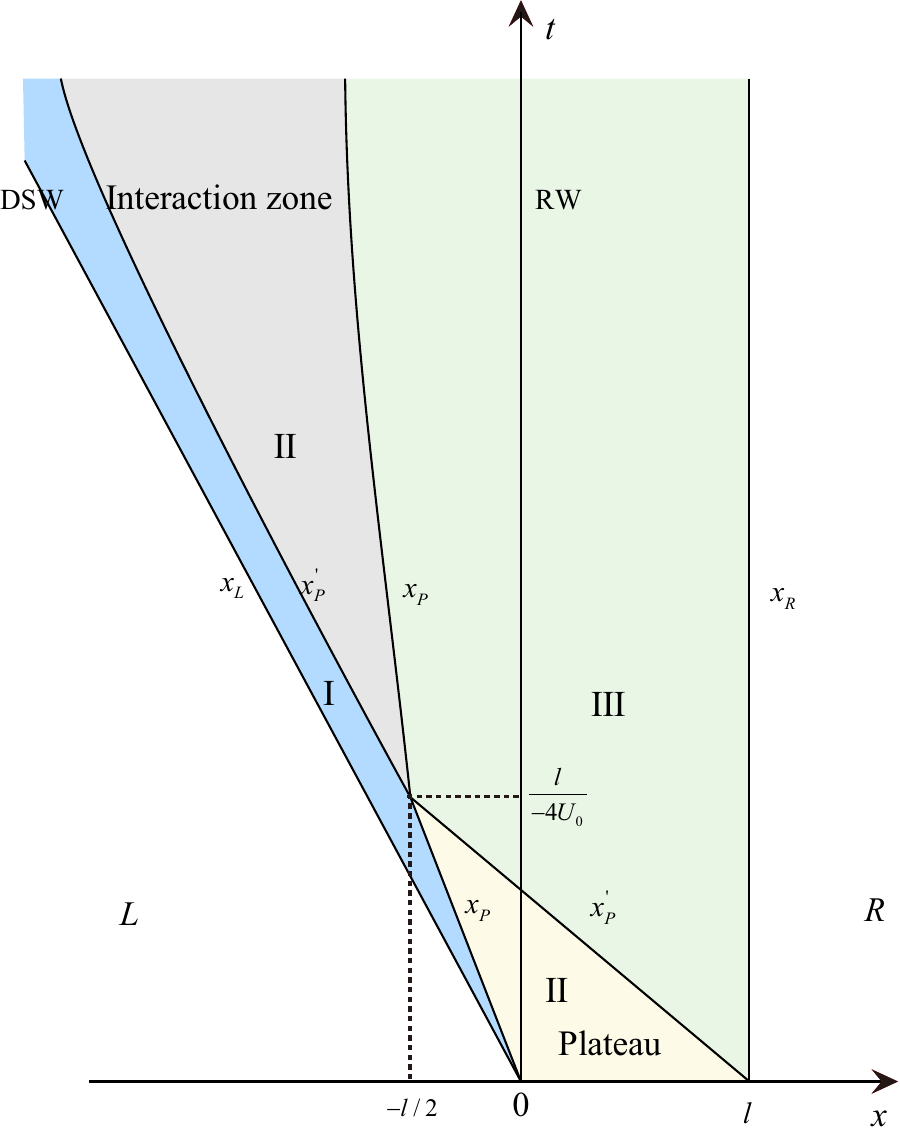}
\caption{{\protect\small Dynamical evolution of the well type initial value without trial soliton in the $(x,t)$ plane.}}
\label{well-whitham-fig}
\end{figure}

The initial potential of the step up leads to the appearance of RW, while the step down corresponds to DSW, which is the classic conclusion of the Gurevich-Pitaevskii problem \cite{Gurevich-Pitaevskii1973} of the KdV equation. The direct influence of the two discontinuous of the initial value is the emergence of interaction zone \cite{EL-Chaos-2002}. The evolution of the well initial value is numerically simulated, see Fig. \ref{lambda-whitham-fig}, where five regions are presented in a short time, with DSW region, plateau region, and RW region between the zero background regions on both sides. After the critical moment occurs, the plateau region is replaced by the modulated linear wave region. Fig. \ref{well-whitham-fig} shows the evolution of the well initial value in the $(x,t)$ plane, The labels of the boundaries and regions are shown in the figure and are used in subsequent notations. To describe this process, it is necessary and natural to focus on the boundaries of the each region.
\par
Before interactions occur, the calculation of the boundary is straightforward, that is, in finite time (can be calculated exactly), the plane is divided into five regions, and the three plane wave regions sandwich the DSW and RW regions. The boundaries are as follows:
\begin{equation}
\begin{aligned}
\label{before-bound}
&x_L=v_2(U_0,U0,0)t=12U_0t,~
x_P=v_2(U_0,0,0)t=2U_0t,\\
&x_P^{'}=l+v_1(U_0,0,0)t=l+6U_0t,~
x_R=l+v_1(0,0,0)t=l.
\end{aligned}
\end{equation}
\par
The critical moment corresponds to the overlap of the two boundaries, that is, $x_P=x_P^{'}$. We mark this moment as $t^*=\frac{l}{-4U_0}$.
\par
The initial value considered in this paper leads to $\lambda_3=0$, focusing on such a $2\times2$ Whitham system
\begin{equation}
\label{whitham-2}
\begin{aligned}
&\frac{\partial \lambda_1}{\partial t}+v_1(\lambda_1,\lambda_2,0)\frac{\partial \lambda_1}{\partial x}=0,\\
&\frac{\partial \lambda_2}{\partial t}+v_1(\lambda_1,\lambda_2,0)\frac{\partial \lambda_2}{\partial x}=0.\\
\end{aligned}
\end{equation}
Introducing Hodograph transformation $x-v_jt=w_j$, rewriting (\ref{whitham-2}) as
\begin{equation}
\label{hodograph-system}
\frac{\partial_{1} w_2}{w_1-w_2}=\frac{\partial_{1} v_2}{v_1-v_2},\quad
\frac{\partial_{2} w_2}{w_2-w_1}=\frac{\partial_{2} v_2}{v_2-v_1},
\end{equation}
where $\partial_{\lambda_j}=\partial_{j}$ and $w_j=w_j(\lambda_1,\lambda_2)$ that can be expressed by an unknown function $f=f(\lambda_1, \lambda_2)$ as
\begin{equation}
\label{wj}
w_j=\frac{\partial_j(kf)}{\partial_j k}=f-\frac{L}{\partial_j L}\partial_j f, \quad j=1,2.
\end{equation}
Arrange (\ref{hodograph-system}) and (\ref{wj}) as
$$
2(\lambda_1-\lambda_2)\partial_{23}^2f=\partial_1 f-\partial_2 f.
$$
which is an Euler-Darboux-Poisson equation that allows a solution of the form \cite{Tricomi-1961}
$$
f(\lambda_1,\lambda_2)=l-\frac{l}{\pi}\int_{\lambda_2}^0\frac{\sqrt{U_0-\beta} d \beta}{\sqrt{\beta(\beta-\lambda_2)(\beta-\lambda_1)}}.
$$
It can be expressed explicitly by using elliptic integrals.
\par
For $x_P$, this is the boundary connecting the interaction region to the RW, and is essentially controlled by the Riemann invariant $\lambda_1$ that satisfies the Hopf equation. We have
\begin{equation}
\label{x_P}
\frac{d x_P}{d t}=v_2(\lambda_1,0,0)=v_3(\lambda_1,0,0)=2\lambda_1.
\end{equation}
Taking $t^*=-\frac{l}{4U_0}, x^*=-\frac{l}{2}$ as the condition for determining the solution, we can get
$$
x_P=l-\frac{3}{2}(\sqrt{-4U_0}l)^{\frac{2}{3}}t^{\frac{1}{3}}.
$$
The boundary $x_P$ connecting the interaction region and the dispersive shock wave region is
$$
\begin{aligned}
&t_P^{'}=(1+\frac{(1-m)(1-\mu(m))}{(2-m)\mu(m)-2(1-m)})\frac{l}{4U_0\sqrt{m}},\\
&x_P^{'}=-(1+\frac{3m(1-m)}{(2-m)\mu(m)-2(1-m)})\frac{l}{2\sqrt{m}},
\end{aligned}
$$
where $\mu(m)=\frac{E(m)}{K(m)}$ and $m=-\frac{\lambda_2-U0}{U_0}$.
\par
According to $t=\frac{w_1-w_2}{v_2-v_1}$, note that when $t\rightarrow \infty$, one has $m \rightarrow 0$, which means that $a=2(\lambda_2-\lambda_1)$ gradually approaches $0$, i.e., the oscillations in the interaction zone asymptotically convert into the linear wave train.

\section{Solitonic modulation system}\label{sec:3}

\subsection{Riemann invariants}

Consider the general form of one-dimensional dispersive hydrodynamics, which is a hyperbolic conservation law model modified by the dispersion term, (see \cite{EL-JFM-2021},\cite{EL-PRL-2018})
$$u_t+F(u)_x=(D[u])_x,$$
taking the hydrodynamic flux function $F(u)=3u^2$, the term that differential operator acts on $D[u]=-u_{xx}$, then it corresponds to the KdV equation (\ref{KdV}).
\par
The KdV equation (\ref{KdV}) has the following dispersion relationship,
$$
\omega(k,\overline{u})=6\overline{u}k-k^3, \quad k\in \mathbb{R},\quad k\neq 0.
$$
which connects the local wave number $k$ with the local frequency $\omega$ and determines the kinematic properties of the wave train. In modulation theory, if the phase is labeled by $\theta=kx-\omega t$, then we have
$
k(x,t)=\frac{\partial \theta}{\partial x}, \quad \omega(x,t)=-\frac{\partial \theta}{\partial t}.
$
The wave number $k$ and frequency $\omega$ of the modulation system are combined to form the conservation equation, i.e.,
$$k_t+\omega_x=0.$$
Solitonic dispersive hydrodynamics could be described analytically based on the soliton reduction of Whitham modulation equation. The modulation system is expressed by physical wave parameters of the form
$$
\mathbf{u}_t+\mathbf{V}(\mathbf{u})\mathbf{u}_x=0,
$$
where for the KdV equation $\mathbf{u}\in \{(\overline{u},a,k)|\overline{u}\in \mathbb{R}, a>0, k \in \mathbb{R}\backslash \{0\} \}$ and the mean flow $\overline{u}$,
amplitude $a$, and wave number $k$ are all slowly varying functions of $x$ and $t$.
For the noninteracting soliton wave train regime, consider the limit state $k\rightarrow 0$, then the modulation system admits the following exact reduction:
\begin{equation}
\begin{aligned}
\label{soliton-modulation-1}
&\overline{u}_t+6\overline{u}\overline{u}_x=0, \\
&a_t+g(a,\overline{u})\overline{u}_x+c(a,\overline{u})a_x=0,
\end{aligned}
\end{equation}
where $g(a,\overline{u})$ is a coupling function related to the background mean flow $\overline{u}$ and soliton amplitude $a$, and $c(a,\overline{u})$ represents the soliton amplitude-speed relation.
According to the conservation of waves, the equation below describes the soliton train
\begin{equation}
\label{soliton-modulation-2}
k_t+(c(a,\overline{u})k)_x=0,
\end{equation}
Formula (\ref{soliton-modulation-1}) and formula (\ref{soliton-modulation-2}) constitute the solitonic modulation system.
\par
System (\ref{soliton-modulation-1}) can be rewritten into a diagonal form by defining $q=gd\overline{u}+(c-6\overline{u})da$
\begin{equation}
\label{modulation}
\left(
  \begin{array}{c}
    \overline{u} \\
    q \\
  \end{array}
\right)_t+\left(
            \begin{array}{cc}
              6\overline{u} & 0 \\
              0 & C(q,\overline{u}) \\
            \end{array}
          \right)\left(
  \begin{array}{c}
    \overline{u} \\
    q \\
  \end{array}
\right)_x=\left(
  \begin{array}{c}
    0 \\
    0 \\
  \end{array}
\right),
\end{equation}
where $C(q,\overline{u})\equiv c(q,\overline{u})$. The KdV equation is strictly hyperbolic, that is, the dispersive hydrodynamics convex, which requires $C\neq 6\overline{u}$.
It is pointed out here that $\overline{u}$ is ``hydrodynamic" Riemann invariant and satisfies the initial value condition, while $q$ is the ``solitonic" Riemann invariant, which exists as an adiabatic invariant \cite{EL-PRL-2018}.
\par
Introduce a conjugate (soliton) wave number $\widetilde{k}$, satisfying
$c(a,\overline{u})=\frac{\widetilde{w}(\widetilde{k},\overline{u})}{\widetilde{k}},$
where $\widetilde{w}(\widetilde{k},\overline{u})=-iw(i\widetilde{k},\overline{u})$. Then, we have
$\widetilde{k}^2=2a$, which implies that $\widetilde{k}\rightarrow 0$ is equivalent to $a\rightarrow 0$, i.e., $q=4\overline{u}$. This means that system (\ref{modulation}) degenerates into a single hyperbolic equation $\overline{u}_t+6\overline{u}\overline{u}_x=0$. It happens to be the KdV equation after removing the dispersion term, also known as the Hopf equation.
\par
According to \cite{EL-Chaos-2005}, the adiabatic invariants can be obtained directly from the following equation
$$
\frac{d\widetilde{k}}{d\overline{u}}=
\frac{\widetilde{w}_{\overline{u}}}{6\overline{u}-\widetilde{w}_{\widetilde{k}}}.
$$
After integral operation, it follows
\begin{equation}
\label{q}
q(a,\overline{u})=4\overline{u}+2a,
\end{equation}
and
\begin{equation}
\label{C}
C(q,\overline{u})=2\overline{u}+q.
\end{equation}
In order to rewrite equation (\ref{soliton-modulation-2}) in diagonal form, we define
\begin{equation}
p(q,\overline{u})=\mathrm{exp}(-\int_{\overline{u}_0}^{\overline{u}}\frac{C_u(q,u)}{F^{'}(u)-C(q,u)}du).
\end{equation}
Reminding $C(q,u)=2u+q$ and $F^{'}(u)=6u$, then $p$ is determined by
\begin{equation}
p(q,\overline{u})=\mathrm{exp}\left\{-\int_{\overline{u}_0}^{\overline{u}}\frac{2}{6u-(2u+q)}du\right\}
=\frac{1}{\sqrt{(4\overline{u}-q)}},
\end{equation}
where it is convenient to take $\overline{u}_0=\frac{1-q}{4}$. In fact, if $q$ is a constant, this is also consistent with the soliton-mean flow interaction we consider. The degraded system (\ref{modulation}) combined with equation (\ref{soliton-modulation-2}) still results in a hyperbolic system. In other words, the following equation can be used to replace equation (\ref{soliton-modulation-2}) with
$$
(kp)_t+C(q,\overline{u})(kp)_x=0.
$$
In fact, $r=kp(q,\overline{u})$ is regarded as an adiabatic invariant, which is the third Riemann invariant \cite{EL-PRL-2018}. Notice that the determination of Riemann invariants is very important. The constants of $q$ and $r$ directly lead to the transmission condition. The trajectory of the soliton after interacting with the mean flow also depends on this.

\par
In fact, when $m\rightarrow1$ (i.e., $\lambda_2\rightarrow\lambda_3$), the Whitham modulation system degenerates to
\begin{equation}
\begin{aligned}
&\frac{\partial \lambda_1}{\partial t}+6\lambda_1\frac{\partial \lambda_1}{\partial x}=0, \\
&\frac{\partial \lambda_3}{\partial t}+(2\lambda_1+4\lambda_3)\frac{\partial \lambda_3}{\partial x}=0.
\end{aligned}
\end{equation}
Denote the background wave as $\overline{u}=\lambda_1$ and the soliton amplitude as $a=2(\lambda_3-\lambda_1)$. Then the correspondence with the soliton modulation system can be achieved at
\begin{equation}
\label{re-whitham}
\begin{aligned}
&\frac{\partial \overline{u}}{\partial t}+6\overline{u}\frac{\partial \overline{u}}{\partial x}=0, \\
&\frac{\partial a}{\partial t}+(6\overline{u}+2a)\frac{\partial a}{\partial x}+2a\frac{\partial \overline{u}}{\partial x}=0,
\end{aligned}
\end{equation}
where the coupling function $g(a,\overline{u})=2a$, and $C(\overline{u},a)=6\overline{u}+2a$.
Based on $q=4\overline{u}+2a$, we can get
\begin{equation}
\label{q-soliton}
\frac{\partial q}{\partial t}+(2\overline{u}+q)\frac{\partial q}{\partial x}=0.
\end{equation}
This equations (\ref{re-whitham})-(\ref{q-soliton})  are actually the modulation system (\ref{modulation}).
\par

\begin{figure}
\centering
\includegraphics[width=6.0cm]{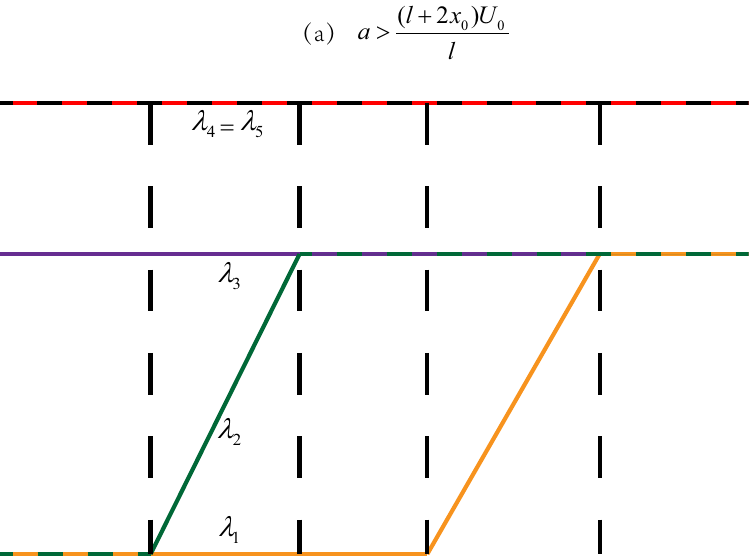}
\includegraphics[width=6.0cm]{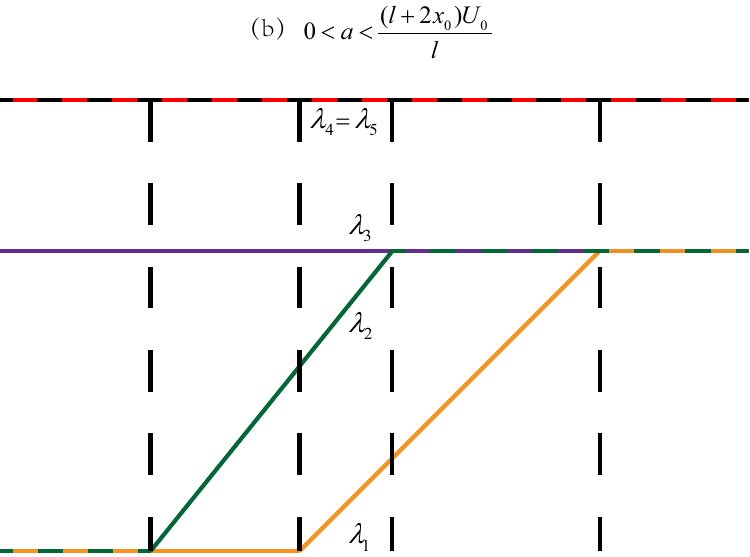}
\includegraphics[width=6.0cm]{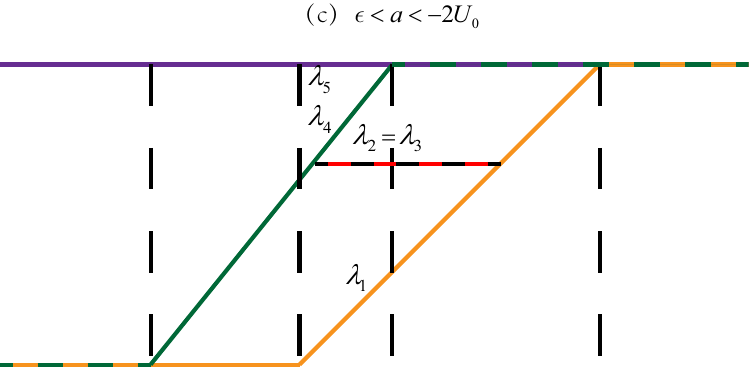}
\includegraphics[width=6.0cm]{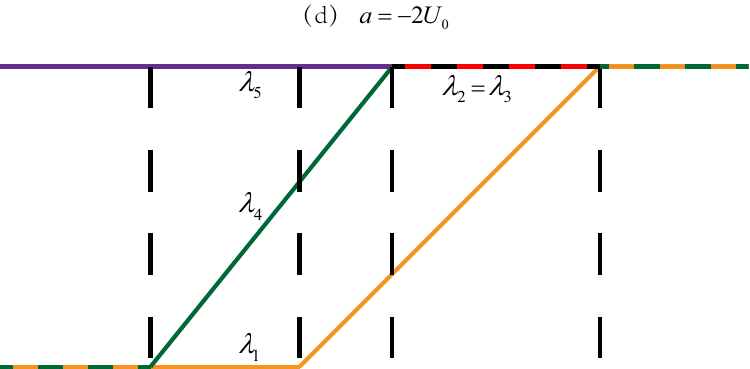}
\includegraphics[width=6.0cm]{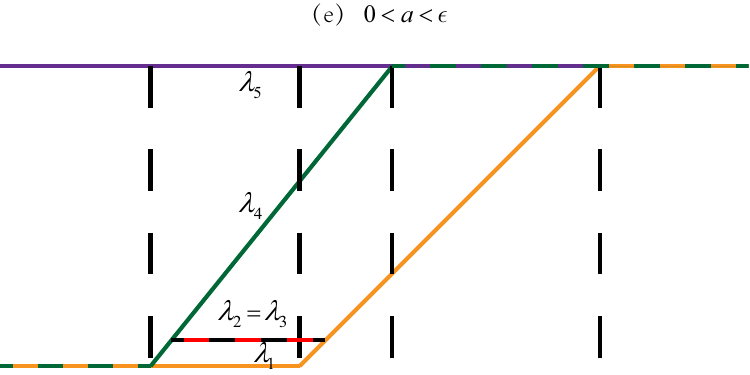}
\caption{{\protect\small Construction of Riemann invariants: (a) Soliton tunnels before the plateau disappears; (b) Soliton tunnels after disappearing in the plateau; (c) Soliton embeds in the RW region; (d) Solito embeds in the LW region; (e) Soliton embeds in the DSW region.}}
\label{whitham-tunnel}
\end{figure}

\subsection{Transmission condition and phase relation}
\par
Considering the soliton solution of the KdV equation (\ref{KdV}) of the form
\begin{equation}
\label{exact-soliton}
u=\overline{u}+a\mathrm{sech}^2\left(\sqrt{\frac{a}{2}}(x-(2a+6\overline{u})t-x_0)\right),
\end{equation}
which is a bright soliton ($a>0$) propagating at a speed of $2a+6\overline{u}$ on the background $\overline{u}$. The soliton amplitude and the background together determine the propagation direction of the soliton. In the framework considered in this paper, the soliton is in a well, which makes it possible for the soliton to move to the left, while it always propagates to the right under the zero background.
\par
The soliton modulation system (\ref{soliton-modulation-1}) and (\ref{soliton-modulation-2}) can be described by a simple wave (to be justified), which means that there is only one Riemann invariant that is not a constant, which is $u$ that satisfies the initial value. That is to say for the three Riemann invariants introduced in the previous section, $\overline{u}$ satisfies the initial condition, and $q(a,\overline{u})$ and $r$ serve as global adiabatic invariants.
\par
Since the region evolved from discontinuity is very complicated, the interaction between the soliton and this region can be expressed through the changes before and after. Corresponding to the initial value problem (\ref{Initial condition}), the initial wave field is described by
\begin{equation}
a(x,0)=\left\{
\begin{aligned}
&a_L,  &x\leq 0,\\
&a_M,  &0<x<l,\\
&a_R,  &x\geq l,
\end{aligned}
\right.
~
k(x,0)=\left\{
\begin{aligned}
&k_L,  &x\leq 0,\\
&k_M,  &0<x<l,\\
&k_R,  &x\geq l.\\
\end{aligned}
\right.
\end{equation}
\par
The invariance of the Riemann invariant $q,r$ directly derives the transmission condition and phase relation of the soliton mean flow problem below
\begin{equation}
q(a_L,u_L)=q(a_R,u_R),
\label{transmission}
\end{equation}
\begin{equation}
\label{invariance}
k_Lp(q_0,u_L)=k_Rp(q_0,u_R).
\end{equation}
The phase shift before$(-)$ and after$(+)$ the interaction between the trial soliton and the mean field is denoted as $\Delta x=x_+-x_-$, and satisfies
\begin{equation}
\frac{x_+}{x_-}=\frac{k_-}{k_+}.
\end{equation}
There is a fact that needs to be explained. The step initial data will generate RW and DSW, or the mean flow depends on the order of $u_L$ and $u_R$. Hydrodynamic reciprocity guarantees that we can still find a simple wave solution for soliton-DSW modulation, ensuring that the soliton modulation system (\ref{soliton-modulation-1}) and (\ref{soliton-modulation-2}) are still valid. In fact, the time reversibility of the KdV equation guarantees this hydrodynamic reciprocity \cite{EL-PRE-2018}.
\par
\section{Soliton-mean field interaction}\label{sec:4}
The mean field generated by the well initial value leads to the emergence of
DSW region and RW region. In a short period of time, a plateau is sandwiched between
the DSW region and the RW region. After a long time, the platform region in the middle
disappears, which can be fully calculated by the Whitham modulation system, and an interaction
region is generated, which is a modulated LW region. The soliton solution with the form
(\ref{exact-soliton}) as the initial trial soliton implies that the propagation speed of the bright soliton
 is $2a+6\overline{u}$. If the initial trial soliton on a zero-background is considered,
 it propagates to the right. According to the background evolution regions,
 we can know that the initial soliton placed on the right side will not interact,
 which is not what we are concerned about. The soliton is initially located inside
 the rectangular well, and the different soliton amplitudes cause the soliton to embed or
 tunnel after a long time. It is known that the soliton tunneling
 has been well studied both theoretically \cite{Anderson-1994,Assanto-2009} and experimentally
 \cite{Marest-2017}. The soliton initially placed on the left side will not be
 embedded in a certain region after a long time, which means that the soliton tunneling
 can always be completed. In this section, the physical phenomenon,
 such as soliton tunneling or embedding, will be described by Whitham modulation theory in detail.

\begin{figure*}
\centering
\includegraphics[width=15.0cm]{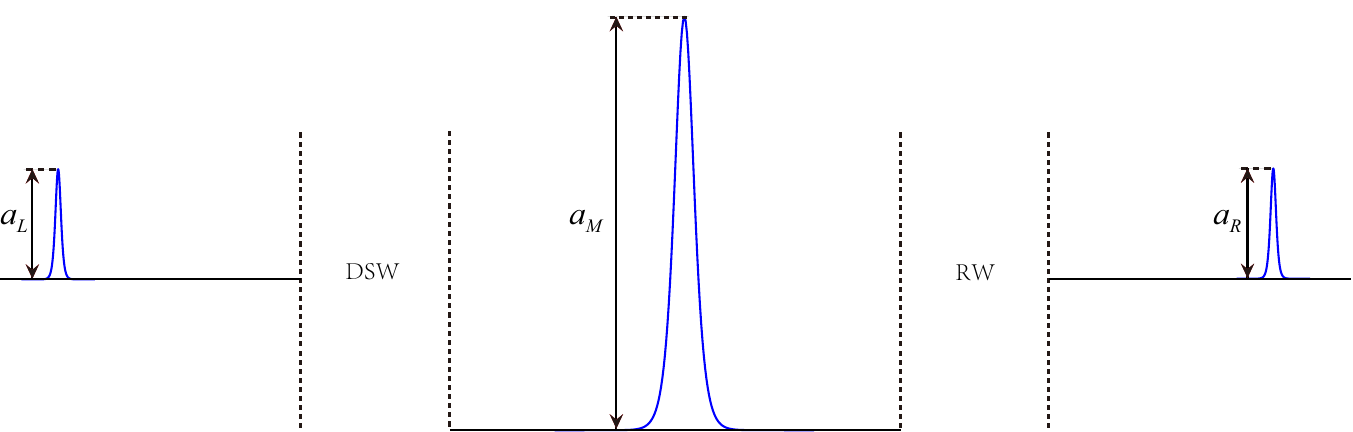}
\caption{{\protect\small When $x_0<0$, the soliton on the left side passes through the DSW region and the RW region successively.}}
\label{tunnel_middle}
\end{figure*}

\subsection{The initial position $x_0<0$}
\par
In the soliton tunneling problem where the initial trial soliton is placed on the left, we take the moment when the interaction region appears as the critical point. This section analyzes the trial soliton separately according to whether it interacts with the modulated LW region. In this case, soliton embedding does not occur. In the framework of Whitham modulation theory, specifically, the finite gap integration method tells us that soliton solution often corresponds to a shrinking of the spectral band. Therefore, for this globally existing soliton solution, a Riemann invariant, $\lambda_{45}$, is constructed to represent the trial soliton. It can be said that this is a degeneration of the KdV-Whitham modulation system of genus two. Fig. \ref{whitham-tunnel}(a) and (b) correspond to the Riemann invariant constructions of the initial trial soliton with initial position $x_0<0$ tunneling before and after the plateau disappears. Initialize the trial soliton approximately as
\begin{equation}
\begin{aligned}
\label{initial soliton1}
u(x,0;x_0)=&\lambda_1-\lambda_2+\lambda_3
+\left(2\lambda_{45}-2(\lambda_1-\lambda_2+\lambda_3)\right)\times\\
&\mathrm{sech}^2(\sqrt{\lambda_{45}-(\lambda_1-\lambda_2+\lambda_3)}(x-x_0)),
\end{aligned}
\end{equation}
where $\lambda_{45}$ is constant, $\lambda_3=0$ and
$$\lambda_1=\left\{
\begin{aligned}
&U_0, x<l\\
&0, x>l,
\end{aligned}
\right. \quad
\lambda_2=\left\{
\begin{aligned}
&U_0, x<0\\
&0, x>0.
\end{aligned}
\right. \quad
$$
According to the two-phase Whitham modulation theory of the KdV equation, $v_{45}$, that is, the soliton velocity $C$, can be calculated in the limit state, i.e.,
\begin{equation}
\begin{aligned}
v_{45} & \equiv {\rm lim}_{\lambda_4\rightarrow\lambda_5}v^{(2)}_4={\rm lim}_{\lambda_4\rightarrow\lambda_5}v^{(2)}_5 ,\\
&=2(\lambda_1+\lambda_2+\lambda_3)+\frac{4(\lambda_{45}-\lambda_2)}
{1-\sqrt{(\frac{(\lambda_{45}-\lambda_2)(\lambda_{3}-\lambda_1)}{(\lambda_{45}-\lambda_3)(\lambda_{45}-\lambda_1)})}}\mathrm{Z}(\psi,m),
\end{aligned}
\end{equation}
where $\mathrm{sin} \psi=\sqrt{\frac{\lambda_{45}-\lambda_3}{\lambda_{45}-\lambda_2}}$ and $\mathrm{Z}(\cdot,\cdot)$ is the Jacobian zeta function.
\par

\begin{table}[htbp]
\footnotesize
\caption{The soliton trajectories in each region}\label{table1}
\begin{center}
  \begin{tabular}{|c|c|} \hline
      \bf Region & \bf Trajectory \\ \hline
     L&$x=x_0+2a_L t$ \\ \hline
     \uppercase\expandafter{\romannumeral1}&$x\sim 2a_Lt+12U_0T_1-2a_LT_1$ \\ \hline
     \uppercase\expandafter{\romannumeral2}& $x=2(a_L+U_0)t-2a_LT_2$\\ \hline
     \uppercase\expandafter{\romannumeral3}& $x=3a_Lt-3(a_L-2U_0)T_3^{\frac{2}{3}}t^{\frac{1}{3}}+l$\\ \hline
     R&$x=2a_Lt-2a_LT_4+l$ \\ \hline
  \end{tabular}
\end{center}
\end{table}
\begin{figure*}
\centering
\includegraphics[width=5.5cm]{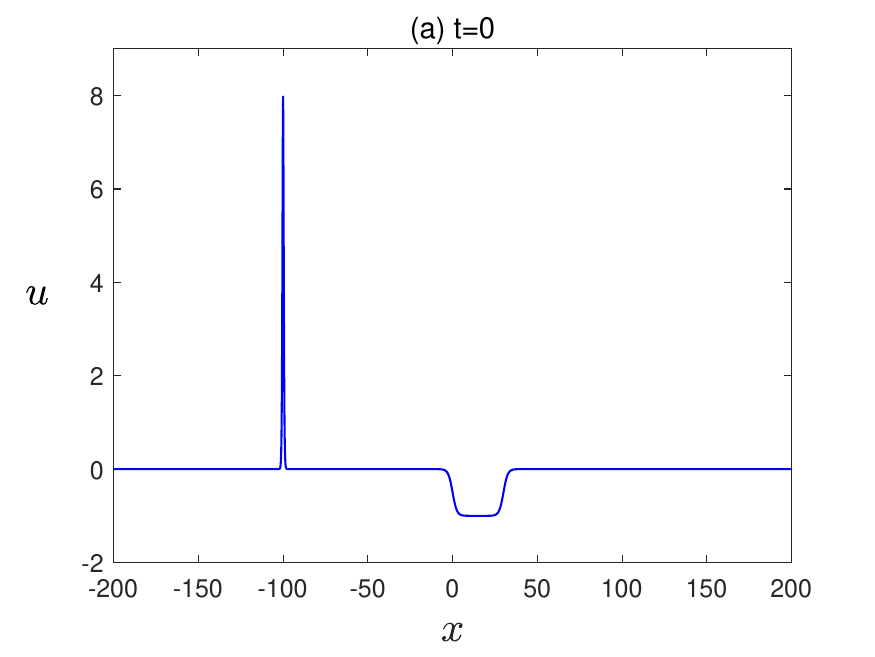}
\includegraphics[width=5.5cm]{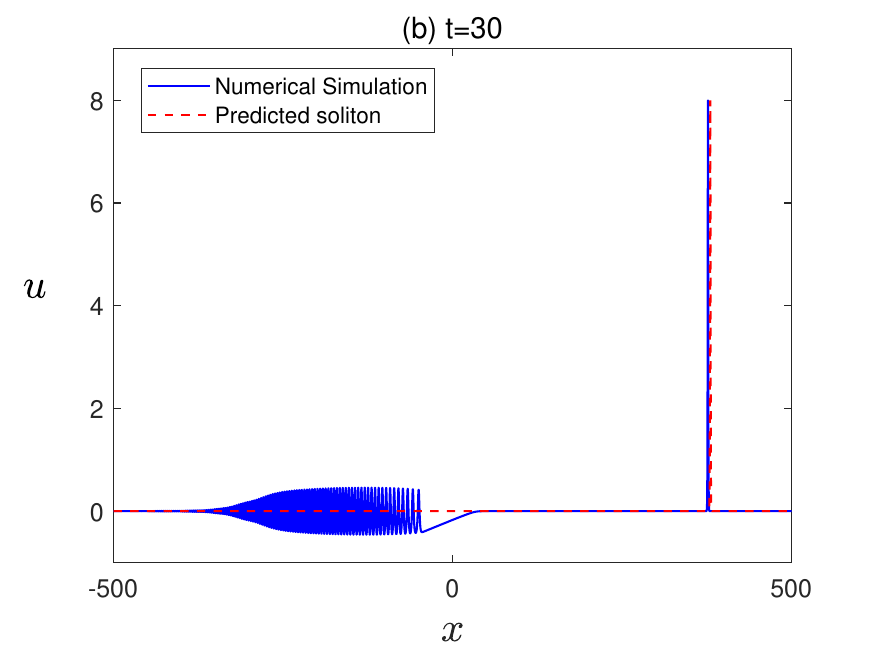}
\includegraphics[width=5.5cm]{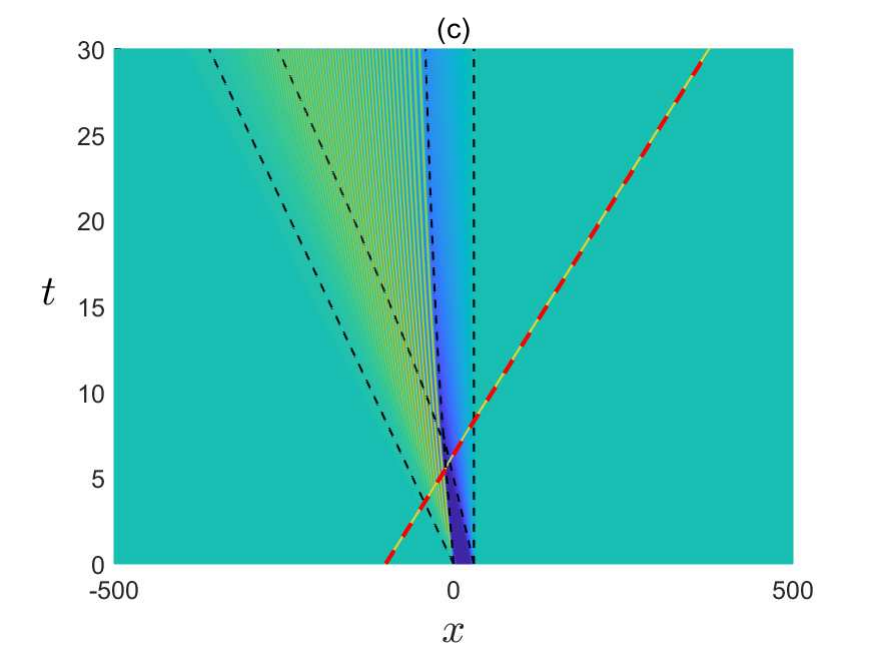}
\caption{{\protect\small (a) Initial value condition; (b) Behavior of soliton interacting with the mean field at $t=30$; (c) The process of soliton interaction with mean field. The black dashed line is the boundary of the region, and the red dashed line is the theoretically predicted soliton trajectory. The initial amplitude of the trial soliton is $a_L=8$, and the initial position is $x_0=-100$.}}
\label{a8}
\end{figure*}
\begin{figure*}
\centering
\includegraphics[width=5.5cm]{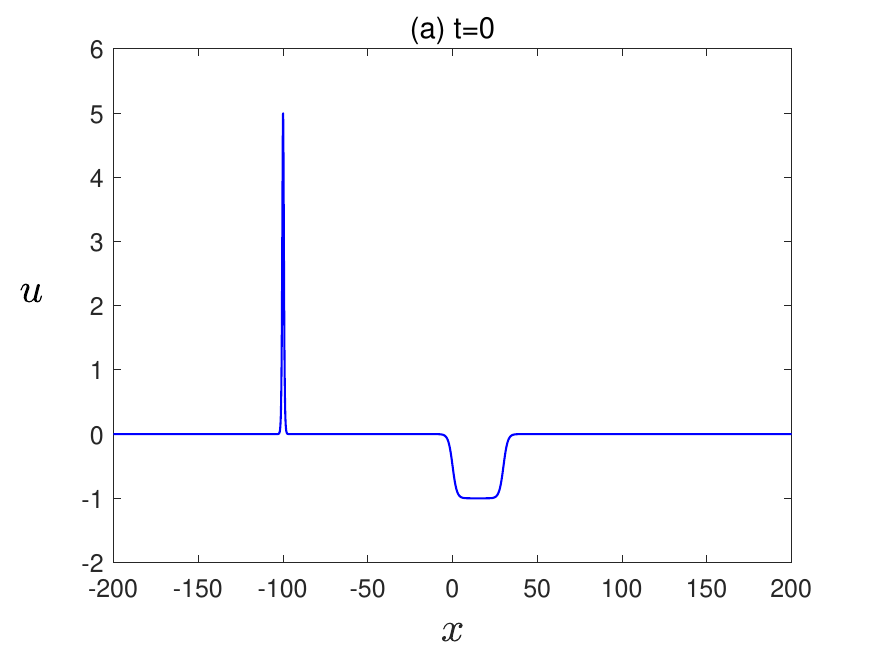}
\includegraphics[width=5.5cm]{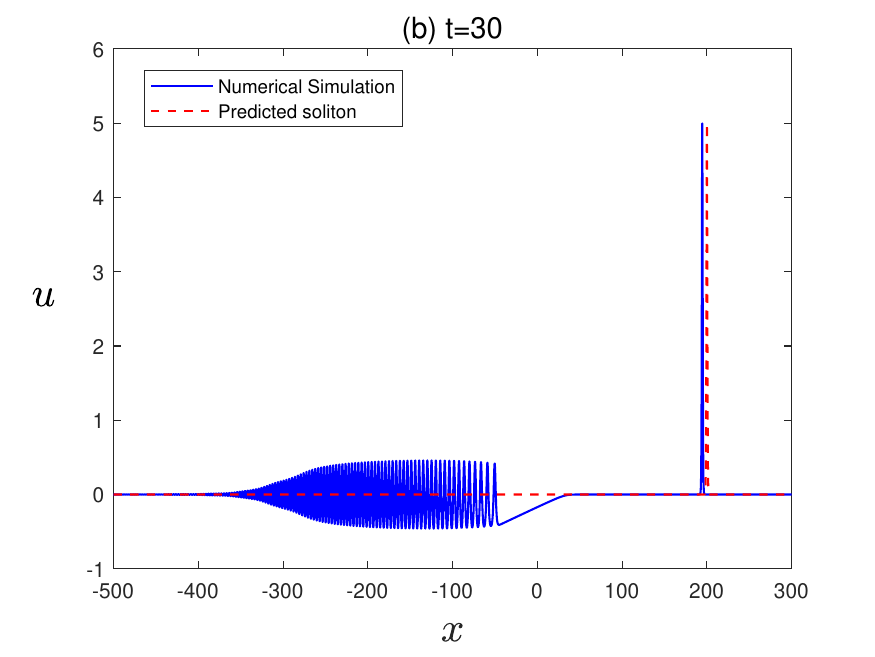}
\includegraphics[width=5.5cm]{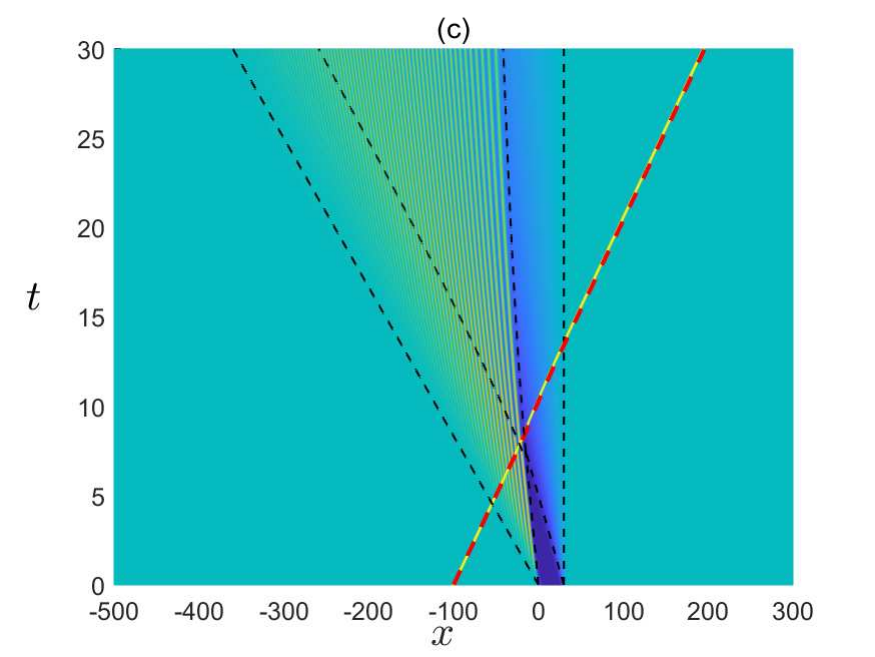}
\caption{{\protect\small (a) Initial value condition; (b) Behavior of soliton interacting with the mean field at $t=30$; (c) The process of soliton interaction with mean field. The black dashed line is the boundary of the region, and the red dashed line is the theoretically predicted soliton trajectory. The initial amplitude of the trial soliton is $a_L=5$, and the initial position is $x_0=-100$.}}
\label{a5}
\end{figure*}

\par
The velocity of a soliton in the mean field is affected by the soliton amplitude and the mean field, and its trajectory satisfies
\begin{equation}
\left\{
\begin{aligned}
&\frac{dx}{dt}=C=6\overline{u}+2a,\\
&x_0=f.
\end{aligned}
\right.
\end{equation}
Here $x_0=f$ is widely considered to be the ``initial" condition of each region, not just $t=0$. Due to the soliton velocity $C=6\overline{u}+2a=q+2\overline{u}$, the soliton trajectory can be obtained in different regions through different solution conditions and background wave conditions.
\par
\begin{itemize}
\item {
{\bf Case \uppercase\expandafter{\romannumeral1}: Before the critical point}}
\end{itemize}
\par
In fact, if the soliton tunneling process does not pass through the interaction region, it will pass through the left plateau region, the DSW region, the middle plateau region, the RW region, and the right plateau region in sequence. This is not difficult to prove. The amplitude of the soliton when it reaches the middle region can be obtained according to (\ref{transmission}), that is
$$a_M=a_L-2U_0.
$$
For the condition $a>-2U_0$ for the soliton to tunnel the rarefaction wave region, this is naturally satisfied, see Fig. \ref{tunnel_middle}.
\par
\par
In the $L$ region in Fig. \ref{well-whitham-fig}, that is, on the left plateau, the soliton propagates stably to the right with velocity $2a_L$ until the boundary of the DSW region, which is marked as $(X_1,T_1)$ on the $(x, t)$ plane and we can directly get $T_1=\frac{-x_0}{2a_L-12U_0}$. From the perspective of Whitham modulation theory, this region corresponds to the limit state of $\lambda_2\rightarrow\lambda_1$, which means that the soliton propagates at a velocity of $v_{45}$ in the mean field of $\overline{u}=\lambda_3$.
\par
In the DSW region, the soliton trajectory can be expressed as follows
\begin{equation}
\left\{
\begin{aligned}
&\frac{dx}{dt}=6u_{\uppercase\expandafter{\romannumeral1}}+2a_{\uppercase\expandafter{\romannumeral1}},\\
&x(T_1)=12U_0T_1.
\end{aligned}
\right.
\end{equation}
According to the solitonic Riemann invariant $q=4\overline{u}+2a$, the soliton velocity can be expressed as $2u_{\uppercase\expandafter{\romannumeral1}}+2a_L$. Here the mean field $u_{\uppercase\expandafter{\romannumeral1}}$ is the modulated DSW. Approximately, the background wave described by DSW is replaced by $u_L$, then $T_2=\frac{6U_0-a_L}{U_0-a_L}T_1$. From the results of numerical simulation, this approximation is feasible.
\par
When the soliton passes through another boundary of the DSW, it enters the next region, which is the plateau region in the middle. The moment when the soliton leaves the plateau \uppercase\expandafter{\romannumeral2} region is marked as $T_3$, and $T_3=\frac{l+2a_LT_2}{2a_L-4U_0}$. In fact, after leaving the DSW region, that is, in the limit state $\lambda_2\rightarrow\lambda_3$, the mean field $\overline{u}=\lambda_1$ is a simple wave RW, which can be written explicitly as
\begin{equation}
\overline{u}(x,t)=\left\{
\begin{aligned}
&U_0,  &2U_0t<x<l+6U_0t,\\
&\frac{x-l}{6t},  &l+6U_0t<x<l,\\
&0,  &l<x.
\end{aligned}
\right.
\label{RW-solution}
\end{equation}
The soliton trajectory is controlled by the following system
\begin{equation}
\label{trajectory-solution3}
\left\{
\begin{aligned}
&\frac{dx}{dt}=6u_{\uppercase\expandafter{\romannumeral3}}+2a_{\uppercase\expandafter{\romannumeral3}}=2a_L+\frac{x}{3t},\\
&x(T_3)=l+6U_0T_3.
\end{aligned}
\right.
\end{equation}
The trajectory of the soliton in the RW region can be expressed. The moment when the soliton passes through the boundary $l$ and leaves the RW region satisfies system (\ref{trajectory-solution3}) and is denoted as $T_4$. Fig. \ref{a8} shows the process of soliton tunneling from the left, and gives the predicted position of the soliton. From the comparison between numerical and theoretical results, it can be seen that the theoretical analysis and approximation are reasonable. Moreover, the soliton trajectories in each region are listed in Table \ref{table1}.

\begin{figure}
\centering
\includegraphics[width=7.5cm]{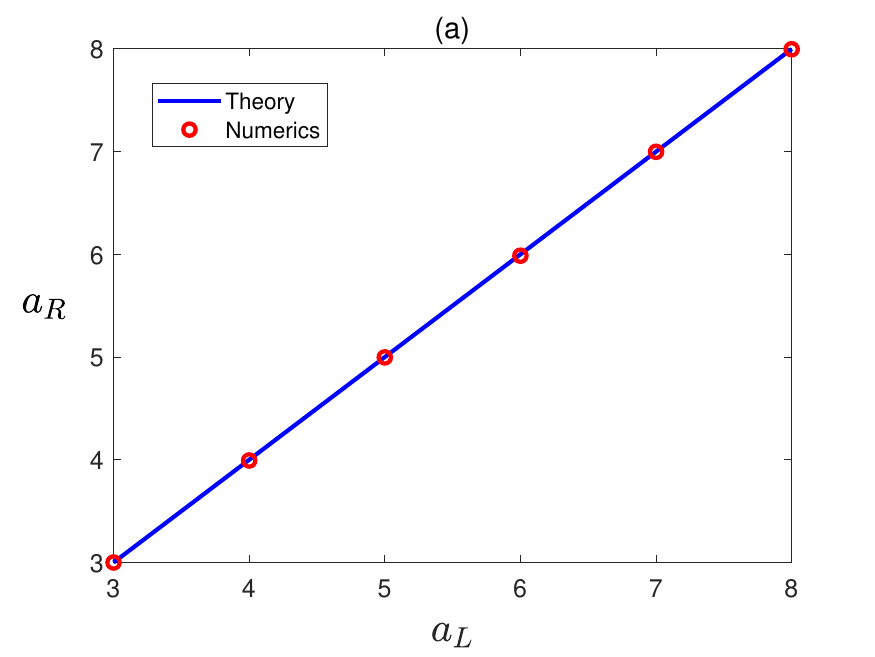}
\includegraphics[width=7.5cm]{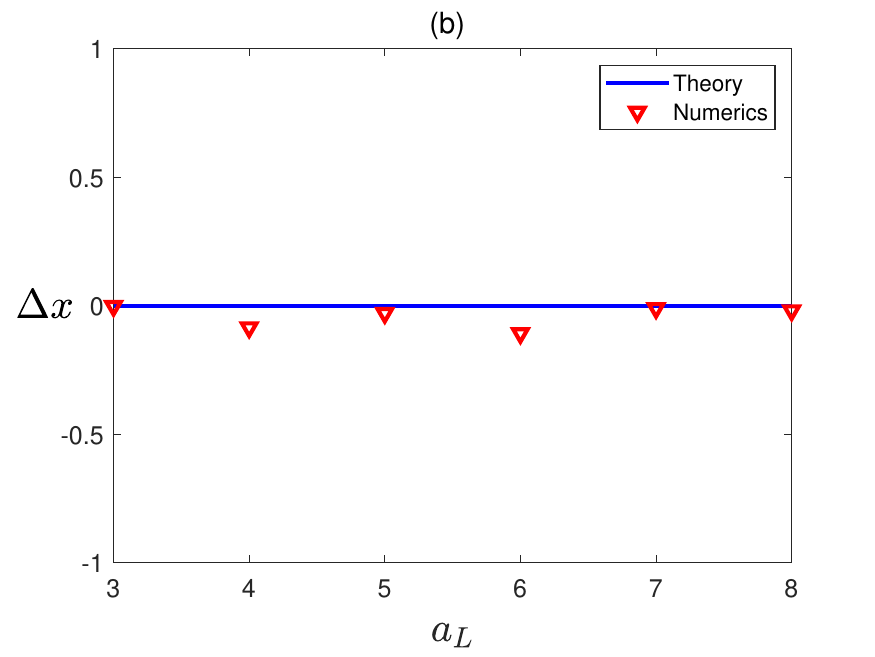}
\caption{{\protect\small Comparison of theoretical predictions of soliton amplitude and phase shift with numerical results for $x_0<0$.}}
\label{pre-num}
\end{figure}

\par
\begin{itemize}
\item {
{\bf Case \uppercase\expandafter{\romannumeral2}: After the critical point}}
\end{itemize}
\par
If the initial velocity of the soliton is small, that is, the initial amplitude of the soliton is small, the soliton will tunnel through the LW region. Since the process of soliton passing through the DSW region cannot be accurately described, the direct impact is that this critical condition can only be given approximately $a_L=\frac{(l+2x_0)U_0}{l}$.
\par
Typically, a single soliton is stable against perturbations that do not cause the soliton to resonantly interact with small-amplitude linear waves. To put it bluntly, soliton should not have consistent phase velocity with linear waves \cite{Buryak-PRE}. In fact, the phase velocity of the linear wave of the KdV equation can be calculated, that is, $6\overline{u}-k^2$, which is wave number dependent and does not coincide with the phase velocity of the soliton in the framework considered in this paper. This ensures that the trial soliton will not radiate after propagating through the LW region.
\par
In this case, we focus on the analysis of soliton tunneling in the LW region. The solution of the modulated linear wave train satisfies (\ref{mean}). In the average sense, when $m\rightarrow 0$ the linear wave train has following form
$$
\left<u\right>=\lambda_2-\lambda_1+\lambda_3,
$$
where the initial value considered in this paper requires $\lambda_3=0$. And $m\rightarrow 0$ if and only if $\lambda_2\rightarrow\lambda_1$, which implies $\left<u\right>\rightarrow 0$. Based on this, the propagation of soliton in the LW region can be approximately expressed as the propagation of soliton under a zero background.
Fig. \ref{a5} shows the dynamic behavior of a soliton passing through the DSW region, LW region, RW region and then tunneling.
\par
For both cases of soliton tunneling, according to (\ref{transmission}) and (\ref{invariance}), the amplitude and phase shift after soliton tunneling can be given
$$
a_R=a_L,\quad \frac{k_L}{k_R}=1,
$$
which implies that neither the amplitude nor the phase has changed. The predicted soliton amplitude is compared with the numerical result in Fig. \ref{pre-num}(a), and both the soliton tunneling before and after the critical point show strong consistency. The predicted and numerical results of the phase shift are shown in Fig. \ref{pre-num}(b).

\begin{figure*}
\centering
\includegraphics[width=5.5cm]{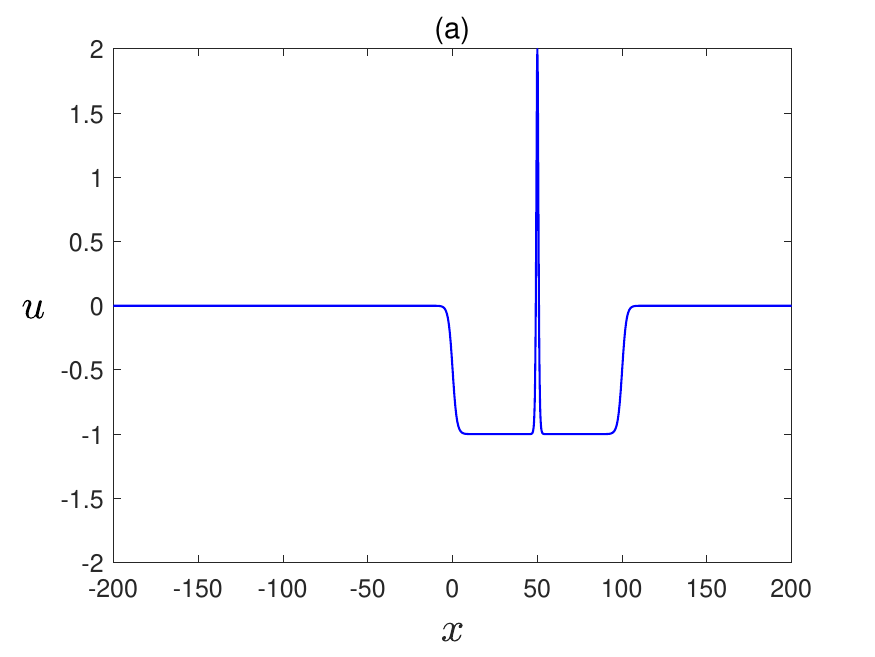}
\includegraphics[width=5.5cm]{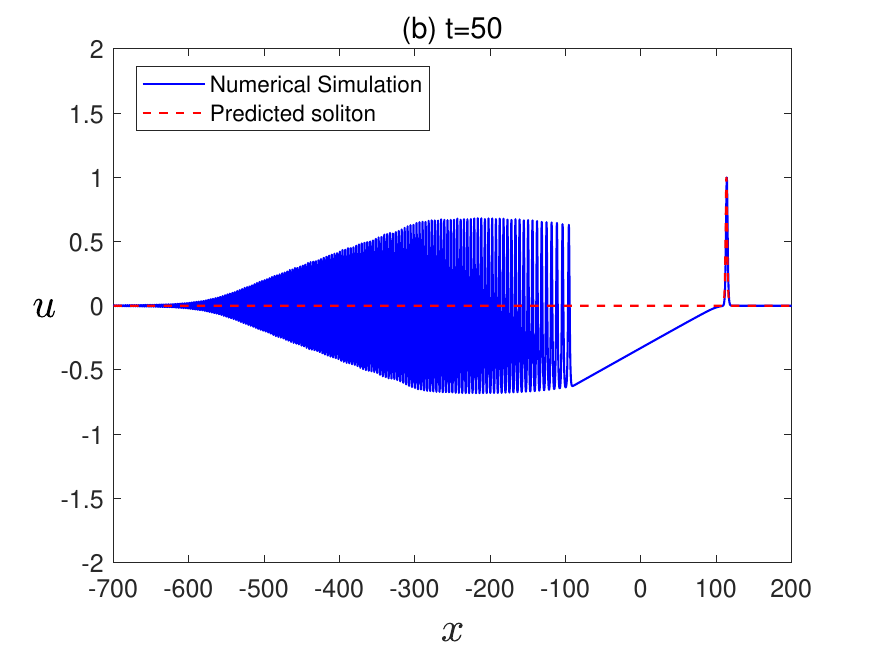}
\includegraphics[width=5.5cm]{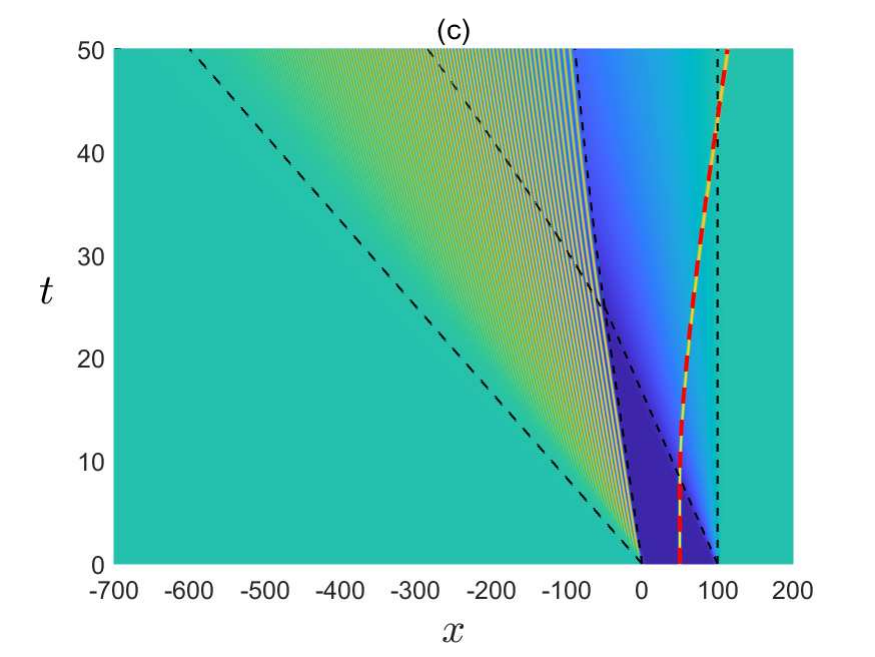}
\caption{{\protect\small (a) Initial value condition; (b) Behavior of soliton interacting with the mean field at $t=50$; (c) The process of soliton interaction with mean field. The black dashed line is the boundary of the region, and the red dashed line is the theoretically predicted soliton trajectory. The initial amplitude of the trial soliton is $a_M=3$, and the initial position is $x_0=50$.}}
\label{a3}
\end{figure*}

\subsection{The initial position $0<x_0<l$}

This subsection considers the initial trial soliton placed in a well, i.e., the initial position of the trial soliton satisfies $0<x_0<l$. The long-term evolution of the soliton exhibits four behaviors: soliton tunneling, soliton embedding into the RW region, soliton embedding into the LW region and soliton embedding into the DSW region.

\subsubsection{Soliton tunneling}
Theoretically, according to (\ref{transmission}), the tunneling condition of soliton can be clearly defined as $a_M>-2U_0$. Fig. \ref{a3} shows the whole process of the trial soliton located in the potential well entering the RW region and then tunneling. According to RW (\ref{RW-solution}), the trajectory of the soliton in the RW region can be obtained
$$
x=-3a_M(t^*)^{\frac{2}{3}}t^{\frac{1}{3}}+(6U_0+3a_M)t+l,
$$
where the trial soliton reaches the RW region at $t^*=\frac{l-x_0}{2a_M}$. According to the transmission condition (\ref{transmission}) and phase condition (\ref{invariance}), the soliton amplitude after tunneling can be obtained as
$$
a_R=a_M+2U_0.
$$
The predicted soliton in Fig. \ref{a3} is in good agreement with the numerical results in terms of both amplitude and position.

\begin{figure*}
\centering
\includegraphics[width=5.5cm]{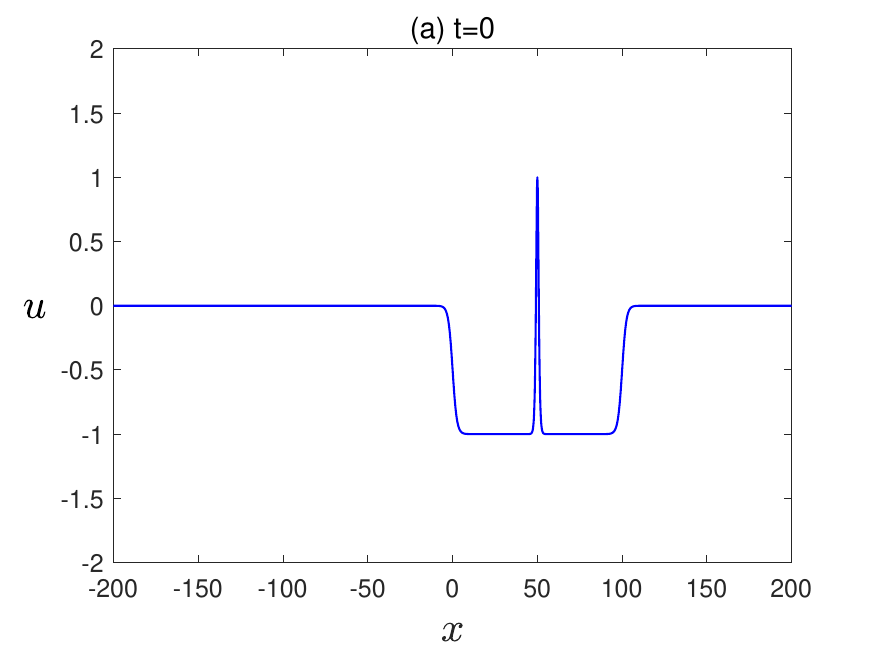}
\includegraphics[width=5.5cm]{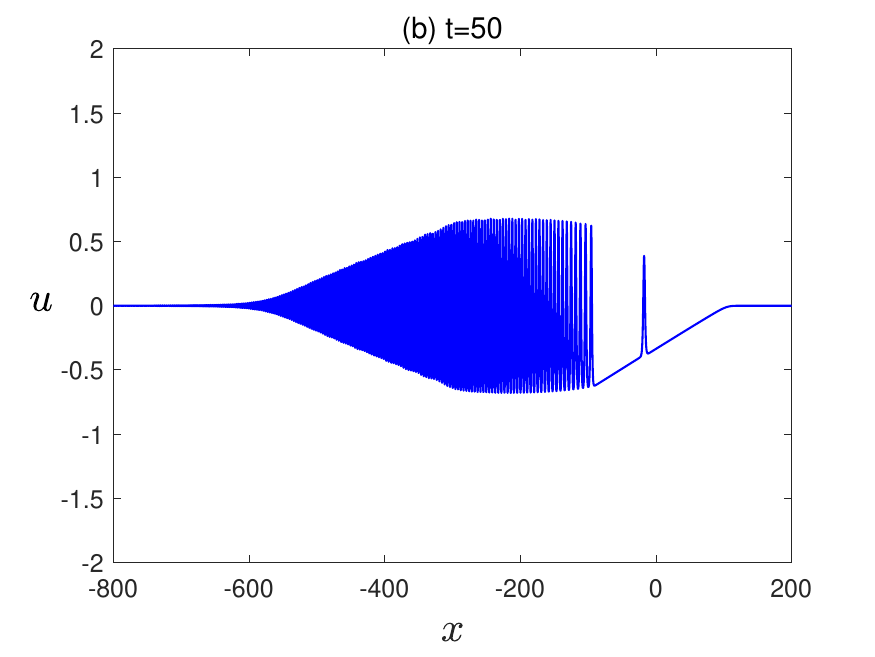}
\includegraphics[width=5.5cm]{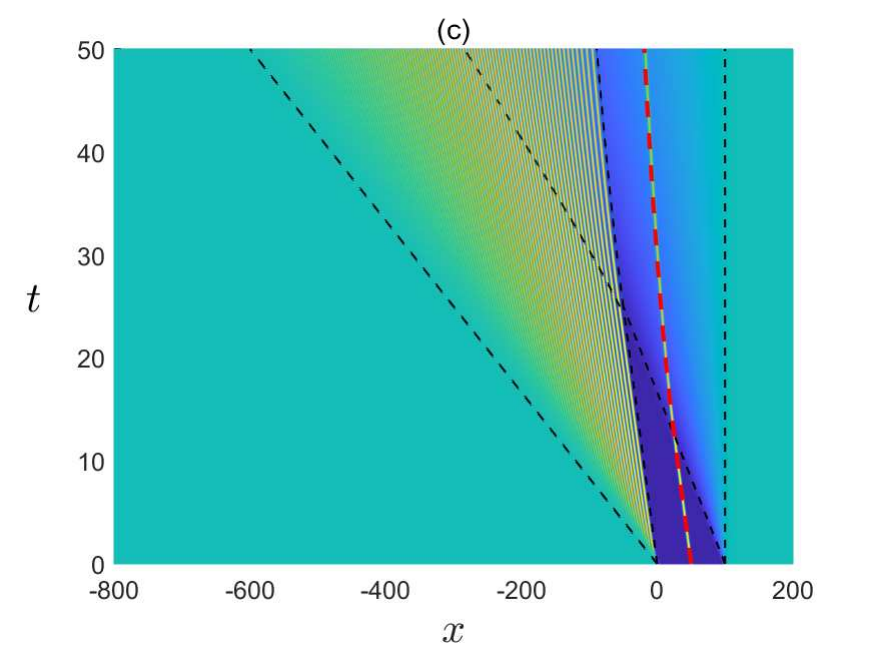}
\caption{{\protect\small (a) Initial value condition; (b) Behavior of soliton interacting with the mean field at $t=50$; (c) The process of soliton interaction with mean field. The black dashed line is the boundary of the region, and the red dashed line is the theoretically predicted trajectory of the trial soliton. The initial amplitude of the trial soliton $a_M=2$, and the initial position is $x_0=50$.}}
\label{a2}
\end{figure*}

\subsubsection{Soliton embedding}
This section considers three embedding cases of $0<a_M\leq-2U_0$, and a detailed analysis is given below.
Approximately initialize the well initial value and modulated soliton by using Riemann invariants in the form
$$
\begin{aligned}
u(x,0;x_0)=&\lambda_1-\lambda_4+\lambda_5+\left(2\lambda_{23}-2(\lambda_1-\lambda_4+\lambda_5)\right)\times\\
&\mathrm{sech}^2(\sqrt{\lambda_{23}-(\lambda_1-\lambda_4+\lambda_5)}(x-x_0)),
\end{aligned}
$$
where $\lambda_5=0$ and
$$\lambda_1=\left\{
\begin{aligned}
&U_0, x<l\\
&0, x>l,
\end{aligned}
\right. \quad
\lambda_4=\left\{
\begin{aligned}
&U_0, x<0\\
&0, x>0.
\end{aligned}
\right. \quad
$$
\par

\begin{figure*}
\centering
\includegraphics[width=5.5cm]{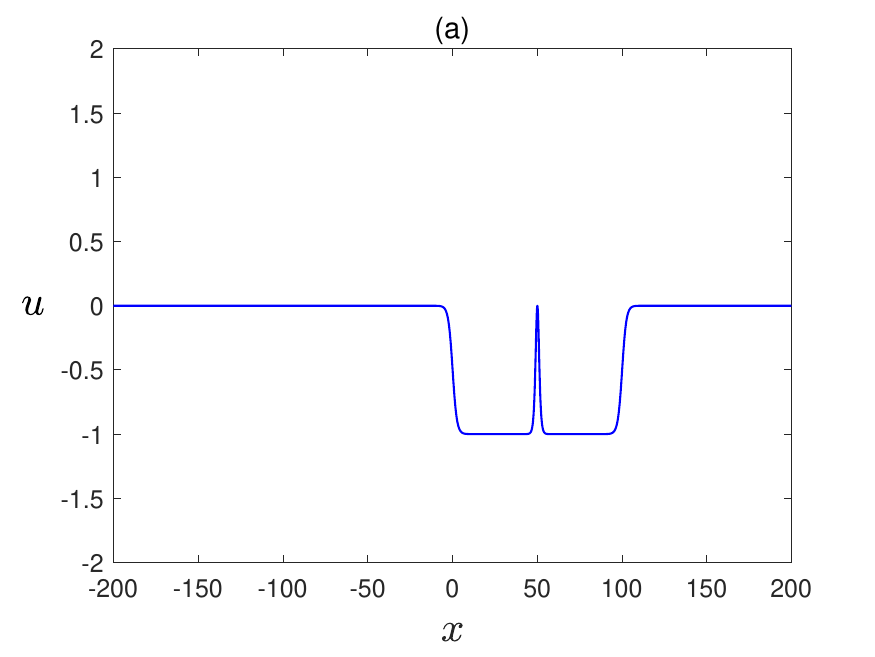}
\includegraphics[width=5.5cm]{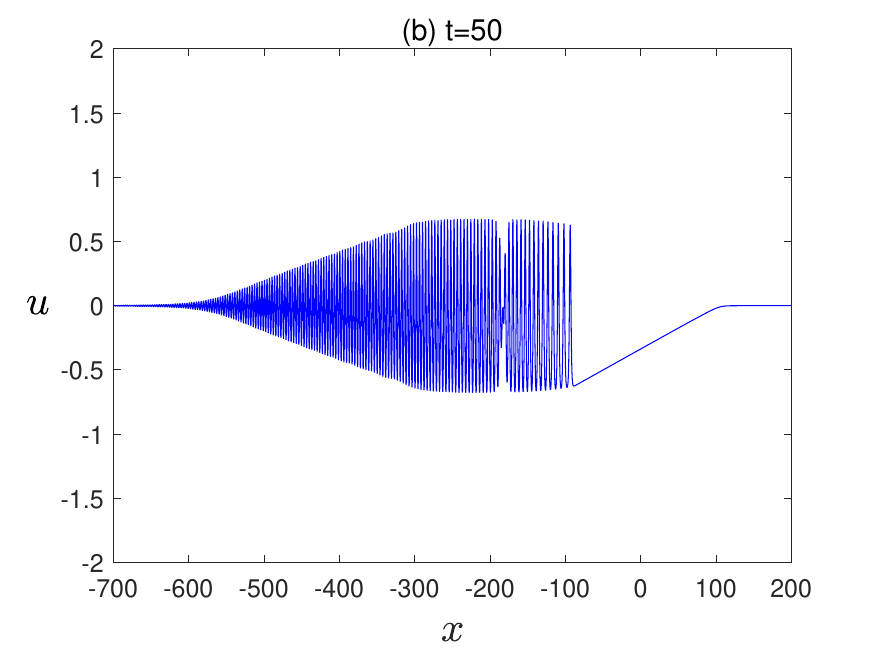}
\includegraphics[width=5.5cm]{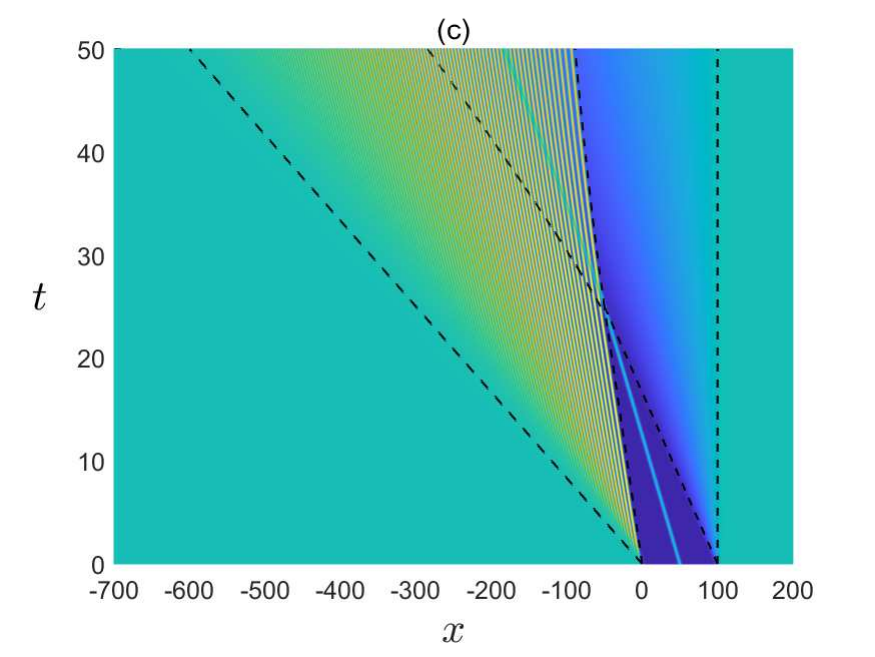}
\caption{{\protect\small (a) Initial value condition; (b) Behavior of soliton interacting with the mean field at $t=50$; (c) The process of soliton interaction with mean field. The black dashed line is the boundary of the region. The initial amplitude of the trial soliton is $a_M=1$, and the initial position is $x_0=50$.}}
\label{a1}
\end{figure*}

\begin{figure*}
\centering
\includegraphics[width=5.5cm]{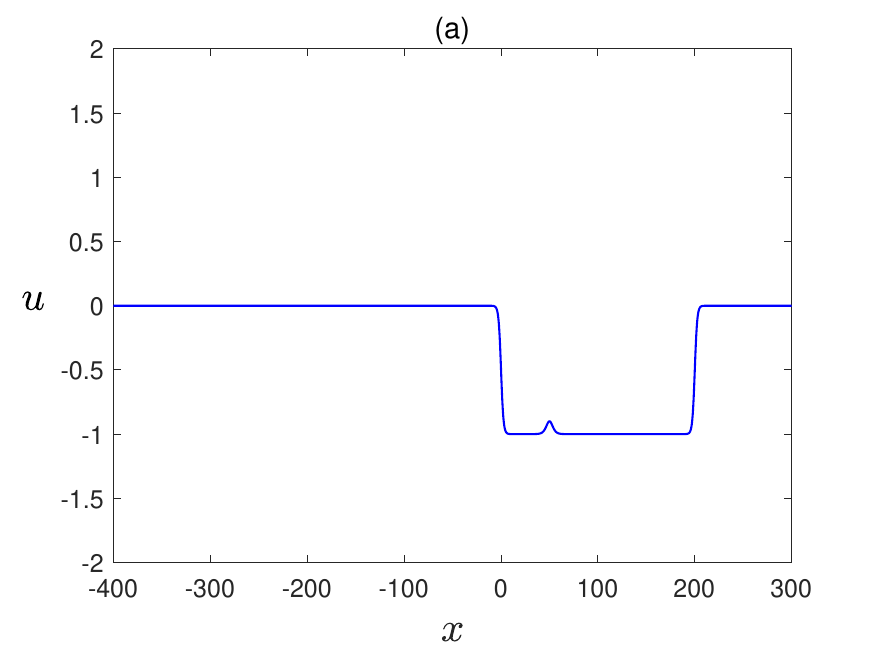}
\includegraphics[width=5.5cm]{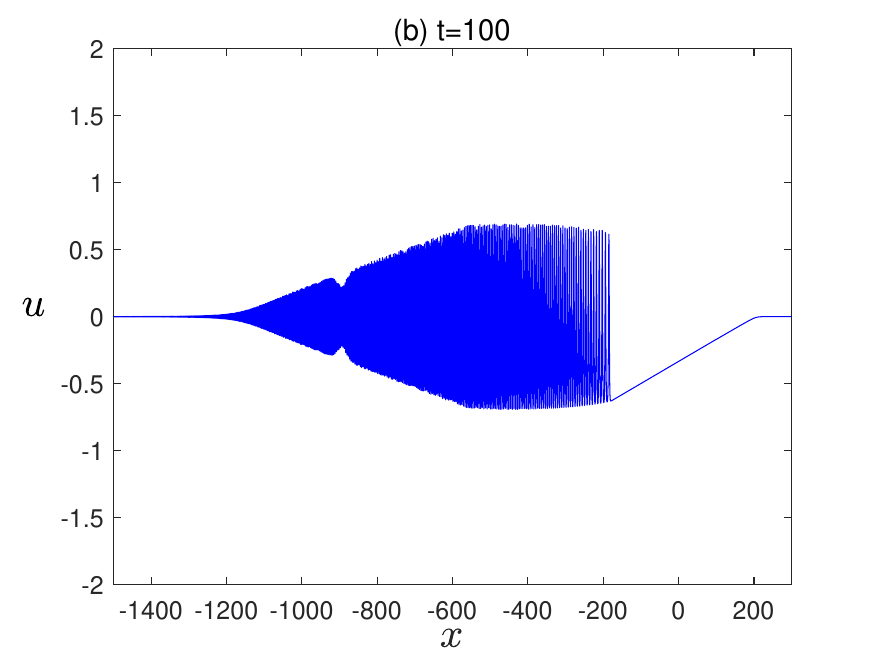}
\includegraphics[width=5.5cm]{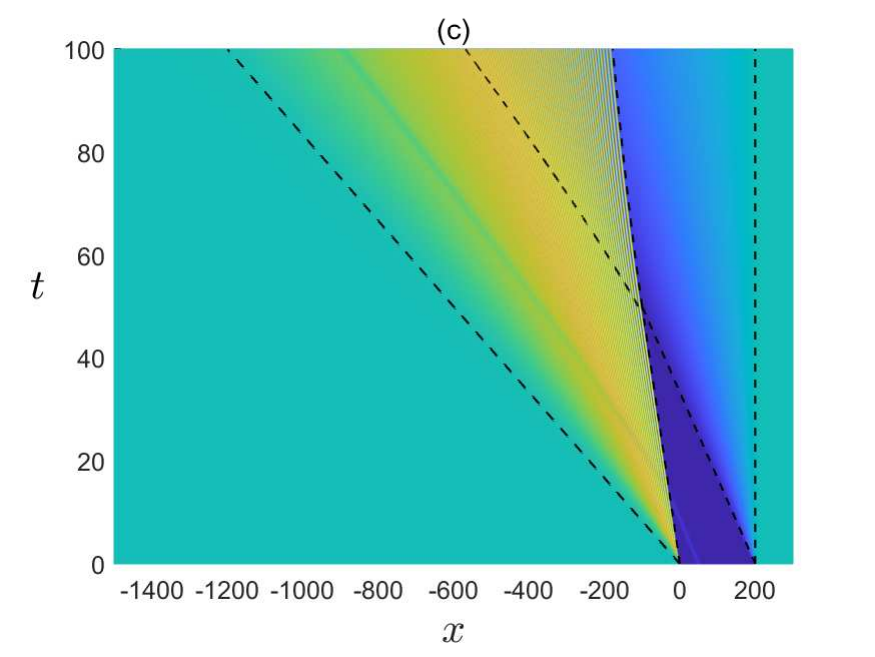}
\caption{{\protect\small (a) Initial value condition; (b) Behavior of soliton interacting with the mean field at $t=50$; (c) The process of soliton interaction with mean field. The black dashed line is the boundary of the region. The initial amplitude of the trial soliton is $a_M=1$, and the initial position is $x_0=50$.}}
\label{a01}
\end{figure*}

\begin{itemize}
\item {
{\bf Case \uppercase\expandafter{\romannumeral1}: soliton embedding into the RW region}}
\end{itemize}
\par
The soliton embedded in the RW region corresponds to the initial soliton amplitude $a_M=-2U_0$ placed in the middle rectangular well, in the mean field of $\overline{u}=\lambda_1$. The distribution of constructing the Riemann invariants is shown in Fig. \ref{whitham-tunnel}(c), where $\lambda_{23}=0$. The soliton velocity can be expressed as
\begin{equation}
\begin{aligned}
v_{23}&=\lim_{\lambda_2\rightarrow \lambda_3}v_2^{(2)}=\lim_{\lambda_2\rightarrow \lambda_3}v_3^{(2)}\\
&=2(\lambda_1+\lambda_4+\lambda_5) -\frac{4(\lambda_5-\lambda_{23})\sqrt{\frac{(\lambda_4-\lambda_{23})(\lambda_{23}-\lambda_1)}{(\lambda_5-\lambda_1)(\lambda_5-\lambda_{23})}}}{\mathrm{Z}(\psi_{23},m_{23})},
\end{aligned}
\end{equation}
where $\mathrm{sin}\psi_{23}=\sqrt{\frac{(\lambda_{23}-\lambda_1)}{\lambda_4-\lambda_1}}$, and $m_{23}=\frac{\lambda_4-\lambda_1}{\lambda_5-\lambda_1}$.
According to the construction of the Riemann invariants in Fig. \ref{whitham-tunnel}(c), the velocity of the embedded soliton can be calculated, corresponding to the limit state $\lambda_4\rightarrow\lambda_{23}$, and the velocity is recorded as $v_e$, that is
$$
v_e=2\lambda_1+4\lambda_{23},
$$
where $\lambda_{23}=0$. This implies that the trial soliton is embedded at a velocity of $2\lambda_1$ after entering the RW region, which is consistent with the boundary velocity of $x_P$(see (\ref{x_P})), once again confirming the conclusion of soliton embedding. The propagation process of the soliton embedded in the RW region satisfies
$$
x=-3a_M(t^*)^{\frac{2}{3}}t^{\frac{1}{3}}+(6U_0+3a_M)t+l.
$$
The time evolution of soliton embedding is shown in Fig.\ref{a2}, in which it is shown that the predicted results of the soliton trajectories are in good agreement with the numerical simulations.
\par
\begin{itemize}
\item {
{\bf Case \uppercase\expandafter{\romannumeral2}: soliton embedding into the LW region}}
\end{itemize}
\par
It is more common for the trial soliton to be embedded in the LW region after long-term evolution, corresponding to $\epsilon < a_M < U_0$, where $0<\epsilon\ll1$. In this case, the distribution of the Riemann invariants is shown in Fig. \ref{whitham-tunnel}(d). When $\lambda_4\rightarrow\lambda_{23}$,  we obtain
$$
v_e=2(\lambda_1+\lambda_{23}+\lambda_5)-4(\lambda_{23}-\lambda_1)\frac{(1-m_e)K(m_e)}{E(m_e)-(1-m_e)K(m_e)},
$$
where $m_e=\frac{\lambda_{23}-\lambda_1}{\lambda_5-\lambda_1}$. This is actually $v_4(\lambda_1,\lambda_{23},\lambda_5)$. Note that when $\lambda_4<\lambda_{23}<0$, due to the strict hyperbolicity of the KdV equation, $v_4<v_{23}<0$, which means that the soliton cannot leave the LW region at a velocity of $v_{23}$. The embedding of the trial soliton is shown in Fig. \ref{a1}.
\par
\begin{itemize}
\item {
{\bf Case \uppercase\expandafter{\romannumeral3}: soliton embedding into the DSW region}}
\end{itemize}
\par
When the initial amplitude of the trial soliton is small enough, it will be embedded in the DSW region after a long period of evolution, as shown in Fig. \ref{a01}. The distribution of the Riemann invariant in this case is shown in Fig. \ref{whitham-tunnel}(e). This embedding mainly depends on the fact that the two boundaries of the DSW region where the soliton front is lost are not contracted for a long time. The soliton embedding in this case is shown in Fig. \ref{a01}.

\subsection{The initial position $x_0>l$}

When the initial position of the trial soliton satisfies $x_0>l$, the rightward propagating trial soliton will not interact with the mean field, which is obvious and not worth discussing further.

\section{Conclusions}\label{sec:5}

This paper focuses on the interaction of trial soliton with the mean field generated by the well type of initial data. The appearance of LW region makes the results more abundant. Different initial positions lead to different propagation directions of the trial soliton, resulting in two results: soliton tunneling and soliton embedding. Specifically, various cases of complete soliton tunneling and soliton embedding in the RW, DSW, and linear wave regions are discussed. The relationship between the initial amplitude of the trial soliton and different results is given in Table \ref{table2}.
\begin{table}[htbp]
\footnotesize
\caption{Conditions for soliton tunneling and soliton embedding}\label{table2}
\begin{center}
  \begin{tabular}{|c|c|c|c|} \hline
  \bf Initial position   & $x_0<0$ & $0<x_0<l$   & $x_0>l$   \\ \hline
    \bf  Tunneling & $a>0$ & $a>-2U_0$ & / \\ \hline
    \bf  Embedding into RW region & / & $a=-2U_0$ &/ \\ \hline
    \bf  Embedding into DSW region & / & $0<a< \epsilon$ & / \\ \hline
    \bf  Embedding into LW region & / & $\epsilon<a<-2U_0$ & / \\ \hline
    \bf No interaction & / & / & $a>0$ \\ \hline
  \end{tabular}
\end{center}
\end{table}
\par
In the framework of Whitham modulation theory, soliton embedding or tunneling is explained from the behavior of Riemann invariants. The trajectories of the trial soliton in different regions are predicted and compared with the numerical results, which match well with each other.

\section*{Acknowledgments}
Support is acknowledged from the National Natural Science Foundation of China, Grant No. 12371247 and No. 12431008.


\par
\bibliographystyle{amsplain}


\end{document}